\documentclass[a4paper,twoside,english,showpacs]{revtex4}
\usepackage{lmodern}

\usepackage[T1]{fontenc}
\usepackage[utf8]{luainputenc}
\usepackage{color}
\usepackage{babel}
\usepackage{amsmath}
\usepackage{amssymb}
\usepackage{graphicx}
\usepackage{esint}
\usepackage[unicode=true,pdfusetitle,
 bookmarks=true,bookmarksnumbered=false,bookmarksopen=false,
 breaklinks=false,pdfborder={0 0 1},backref=false,colorlinks=false]
 {hyperref}

\makeatletter

\pdfpageheight\paperheight
\pdfpagewidth\paperwidth
\@ifundefined{textcolor}{}
{%
 \definecolor{BLACK}{gray}{0}
 \definecolor{WHITE}{gray}{1}
 \definecolor{RED}{rgb}{1,0,0}
 \definecolor{GREEN}{rgb}{0,1,0}
 \definecolor{BLUE}{rgb}{0,0,1}
 \definecolor{CYAN}{cmyk}{1,0,0,0}
 \definecolor{MAGENTA}{cmyk}{0,1,0,0}
 \definecolor{YELLOW}{cmyk}{0,0,1,0}
}

\usepackage{babel}

\usepackage{graphics}
\usepackage{epsfig}
\usepackage{epsf}

\makeatother

\begin{document}

\title{Hartree-Fock Mean-Field Theory for Trapped Dirty Bosons}

\author{Tama Khellil}

\email{khellil.lpth@gmail.com}

\affiliation{Insitute f\"{u}r Theoretische Physik, Freie Universit\"{a}t Berlin,
Arnimallee 14, 14195 Berlin, Germany}

\author{Axel Pelster}

\email{axel.pelster@physik.uni-kl.de}

\affiliation{Fachbereich Physik und Forschungszentrum OPTIMAS, Technische Universit\"{a}t
Kaiserslautern, 67663 Kaiserslautern, Germany}
\begin{abstract}
Here we work out in detail a non-perturbative approach to the dirty
boson problem, which relies on the Hartree-Fock theory and the replica
method. For a weakly interacting Bose gas within a trapped confinement
and a delta-correlated disorder potential at finite temperature, we
determine the underlying free energy. From it we determine via extremization
self-consistency equations for the three components of the particle
density, namely the condensate density, the thermal density, and the
density of fragmented local Bose-Einstein condensates within the respective
minima of the random potential landscape. Solving these self-consistency
equations in one and three dimensions in two other publications has
revealed how these three densities change for increasing disorder
strength.
\end{abstract}


\maketitle

\section{Introduction}

In the dirty boson problem, the combined effect of disorder and two-particle
interaction yields an intriguing interplay between localization and
superfluidity \cite{Intro-11}. Experimentally, the dirty boson problem
was first studied with superfluid helium in porous media like aerosol
glasses (Vycor), where the pores are modeled by statistically distributed
local scatterers \cite{Intro-91,Intro-92,Intro-93,Intro-94}. In Bose
gases disorder appears either naturally as, e.g., in magnetic wire
traps \cite{Intro-75,Intro-76,Intro-77,Fortagh,Schmiedmayer}, where
imperfections of the wire itself can induce local disorder, or it
may be created artificially and controllably as, e.g., by using laser
speckle fields \cite{Intro-78,Intro-79,Intro-21,Goodmann,Intro-16}.
A set-up more in the spirit of condensed matter physics relies on
a Bose gas with impurity atoms of another species trapped in a deep
optical lattice, so the latter represent randomly distributed scatterers
\cite{Intro-13,Intro-14}. Furthermore, an incommensurate optical
lattice can provide a pseudo-random potential for an ultracold Bose
gas \cite{Lewenstein,Ertmer,intro-17}. 

Theoretically, the dirty boson problem can be treated, in principle,
via two complementary approaches. The first one applies the Bogoliubov
theory \cite{Intro-7} and treats disorder, quantum, and thermal fluctuations
perturbatively, which is only valid in systems with sufficiently small
random potential and interaction strength at low enough temperatures
\cite{HM-1}. With this it was found that a weak random disorder potential
leads to a depletion of both the condensate and the superfluid density
due to the localization of bosons in the respective minima of the
random potential. This seminal Huang-Meng theory was later on extended
in different research directions. Results for the shift of the velocity
of sound as well as for its damping due to collisions with the external
field are worked out in Ref.~\cite{HM-3}. Furthermore, the original
special case of a delta-correlated random potential was generalized
to experimentally more realistic disorder correlations with a finite
correlation length, which model, for instance, the pore size dependence
of Vycor glass. A Gaussian correlation was discussed in Ref.~\cite{HM-4},
whereas laser speckles are treated in Refs.~\cite{HM-6,HM-7}. Also
the disorder-induced shift of the critical temperature for the homogeneous
case was analyzed in Refs.~\cite{HM-5,HM-2}, which also has implications
for a harmonic confinement \cite{Timmer}. Furthermore, it was shown
in Refs.~\cite{HM-8,HM-9,HM-10} that dirty dipolar Bose gases yield
even at zero temperature characteristic directional dependences for
thermodynamic quantities due to the anisotropy emerging of superfluidity.
The recent perturbative work \cite{Gaul-Muller-1,Gaul-Muller-2} studies
even in detail the impact of the external random potential upon the
quantum fluctuations. Despite all these many theoretical predictions
of the Huang-Meng theory, which also affect the collective excitations
frequencies of harmonically trapped dirty bosons \cite{Collective-excitations},
so far no experiment has tested them quantitatively.

On the other hand, the dirty boson problem was also tackled non-perturbatively
in different ways. A major result is that increase in the disorder
strength at zero temperature yields a first-order quantum phase transition
from a superfluid to a Bose-glass phase, where in the latter all particles
reside in the respective minima of the random potential. This prediction
is achieved for three dimensions by solving the underlying Gross-Pitaevskii
equation with a random phase approximation \cite{Navez}, as well
as by a stochastic self-consistent mean-field approach using two chemical
potentials, one for the condensate and one for the excited particles
\cite{Yukalov1,Yukalove2}. Dual to that, the non-perturbative approach
of Refs.~\cite{Nattermann,Natterman2} investigates energetically
shape and size of the local minicondensates in the disorder landscape
and deduces from that, for a decreasing disorder strength, when the
Bose-glass phase becomes unstable and goes over into the superfluid.
At finite temperatures the location of superfluid, Bose-glass, and
normal phase in the phase diagram was qualitatively analyzed in Ref.~\cite{Intro-90}
on the basis of a Hartree-Fock mean-field theory with the replica
method. Also Monte-Carlo (MC) simulations have been applied to study
the dirty boson problem. Diffusion MC in Ref.~\cite{Astrakharchik}
obtained the surprising result that a strong enough disorder yields
a superfluid density larger than the condensate density. Furthermore,
worm algorithm MC \cite{Monte-Carlo1,Monte-Carlo2} was able to determine
the dynamic critical exponent of the quantum phase transition from
the Bose-glass to the superfluid in two dimensions.

All those previous theoretical investigations mainly focus
on the possible emergence of the Bose-glass phase and its elusive
properties for homogeneous dirty bosons. Experimentally, however,
ultracold quantum gases have to be confined with the help of a harmonic
trapping potential. Therefore, in case of trapped dirty bosons, there
is a lack of knowledge concerning the Bose-glass region, where the
bosons within the harmonic trap localize in the respective minima
of the superimposed random potential. The present paper works out
in detail a theoretical approach how to describe this localization
of bosons within a harmonic confinement in a systematic way. To this
end we extend the Hartree-Fock mean-field theory of Ref.~\cite{Intro-90}
for a three-dimensional weakly interacting homogeneous Bose gas in
a delta-correlated disorder potential to the experimentally relevant
trapping confinement via a semi-classical approximation and to a general
number of spatial dimensions. By doing so, we work out, in particular,
all the respective technical details which were omitted for brevity
in Ref.~\cite{Intro-90}. In the following we start in Section II
with introducing the functional integral representation of the partition
function for a trapped weakly interacting Bose gas in a disorder potential
at finite temperature. Applying the replica method in Section III
allows to eliminate the random potential right away at the expense
of introducing disorder-induced interactions between different replica
fields, which are nonlocal in both space and time. Then we work out
a Hartree-Fock mean-field theory for this model in Section IV. After
specializing to replica symmetry in Section V, we restrict ourselves
to a delta-correlated disorder potential and contact interaction potential
for the dirty boson model in Section VI. The underlying free energy
is obtained in Section VII. From it we determine via extremization
the underlying self-consistency equations for the three components
of the particle density, namely the condensate density, the thermal
density, and the density of fragmented local Bose-Einstein condensates
within the respective minima of the random potential landscape. The
case of three dimensions is treated in Section VIII, whereas one spatial
dimension is dealt with in Section IX. Note that the two-dimensional
case is not treated in this paper, since our mean-field theory turns
out to diverge in two dimensions, so both a regularization and a subsequent
renormalization is needed, which goes beyond the scope of the present
paper. Furthermore, we introduce the statistical description of a
disorder potential, which is central for describing the dirty boson
problem, as well as the disorder ensemble average in Appendix A. Finally,
Appendix B defines the order parameters for the superfluid and the
Bose-glass phase via off-diagonal long-range order of corresponding
correlation functions.

\section{Bose Model}

We start by considering the model of an $n$-dimensional Bose gas
in an arbitrary trap $V({\bf x})$ and a general interaction potential
$V^{({\rm int})}({\bf x-x'})$ at finite temperature $T$ in $n$
spatial dimensions. The starting point is the functional integral
for the grand-canonical partition function

\begin{equation}
\mathcal{Z}=\mathcal{\oint}\mathcal{D}\psi^{\ast}{\displaystyle \oint}\mathcal{D}\psi e^{-\mathcal{A}\left[\psi^{\ast},\psi\right]/\hslash},\label{eq:Z}
\end{equation}
where the integration is performed over all Bose fields $\psi^{*}({\bf x},\tau),\psi({\bf x},\tau)$
which are periodic in imaginary time $\tau$, i.e.,\textcolor{black}{{}
}$\psi({\bf x},\tau)=\psi({\bf x},\tau+\hbar\beta)$. The Euclidean
action is given in standard notation by 
\begin{eqnarray}
\mathcal{A}\left[\psi^{\ast},\psi\right] & = & \int_{0}^{\hslash\beta}d\tau\int d\mathbf{x}\left\{ \psi^{\ast}\left(\mathbf{x},\tau\right)\left[\hslash\frac{\partial}{\partial\tau}-\frac{\hslash^{2}}{2M}\Delta+V\left(\mathbf{x}\right)+U\left(\mathbf{x}\right)-\mu\right]\psi\left(\mathbf{x},\tau\right)\right.\nonumber \\
 &  & \left.+\frac{1}{2}\int d\mathbf{x}'\psi^{\ast}\left(\mathbf{x},\tau\right)\psi\left(\mathbf{x},\tau\right)V^{({\rm int})}({\bf x-x'})\psi^{\ast}\left(\mathbf{x'},\tau\right)\psi\left(\mathbf{x'},\tau\right)\right\} ,\label{eq:A}
\end{eqnarray}
where $M$ denotes the particle mass, $\mu$ the chemical potential,
$\beta=1/k_{B}T$ the reciprocal temperature, and $T$ the temperature.
Furthermore, $U({\bf x})$ denotes a generally correlated disorder
landscape, whose statistical properties are explained in detail in
Appendix A. 

Note that, in order to guarantee the normal ordering within the functional
integral, we should work with adjoint fields $\psi^{*}({\bf x},\tau^{+})$
with a shifted imaginary time $\tau^{+}=\tau+\eta$ with $\eta\rightarrow0^{+}$
which is infinitesimally later than the imaginary time $\tau$ of
the fields $\psi({\bf x},\tau)$. However, for the sake of simplicity,
we mainly use in the following the notation $\psi^{*}({\bf x},\tau)$
and emphasize the normal ordering only when it is indispensable.

\section{Replica Method}

A standard method to deal with disorder problems is the replica method
\cite{Intro-33,Intro-36,Intro-37}. Instead of treating the actual
problem, one looks at $\mathcal{N}$ copies of the system, then analytically
continues the replicated system to the limit $\mathcal{N}\rightarrow0$.
As the concrete realization of the disorder potential $U({\bf x})$
is not known, the free energy of the system $\Omega$ is defined as
the free energy for fixed disorder potential averaged over all its
realizations 
\begin{equation}
\Omega=-\frac{1}{\beta}\overline{\ln{\cal Z}},\label{F}
\end{equation}
where $\overline{\bullet}$ corresponds to the disorder average over
many realizations. In general it is not possible to explicitly evaluate
expression (\ref{F}), as $\overline{\ln{\cal Z}}\neq\ln\overline{{\cal Z}}\,.$
The replica method is provided by investigating the $\mathcal{N}^{{\rm {th}}}$
power of the grand-canonical partition function ${\cal Z}$ in the
limit $\mathcal{N}\to0$, which yields for the replicated partition
function ${\cal Z}^{\mathcal{N}}=1+\mathcal{N}\,\ln{\cal Z}+\ldots$.
Thus, we deduce for the free energy (\ref{F})

\begin{equation}
\Omega=-\frac{1}{\beta}\,\lim_{\mathcal{N}\to0}\frac{\overline{{\cal Z}^{\mathcal{N}}}-1}{\mathcal{N}}\,.\label{4}
\end{equation}

The fact that all $\mathcal{N}$ replicas are identical simplifies
the calculation further as we will show below. The $\mathcal{N}$-fold
replication of the partition function of the disordered Bose gas in
Eq. \eqref{eq:Z} and a subsequent averaging with respect to the disorder
potential $U({\bf x})$ results in: 
\begin{eqnarray}
\overline{{\cal Z}^{\mathcal{N}}} & = & \left\{ \prod_{\alpha=1}^{\mathcal{N}}\oint{\cal D}\psi_{\alpha}^{*}\oint{\cal D}\psi_{\alpha}\right\} \,\exp\left\{ \frac{-1}{\hbar}\int_{0}^{\hbar\beta}d\tau\int d\mathbf{x}\sum_{\alpha=1}^{\mathcal{N}}\,\left\{ \psi_{\alpha}^{*}({\bf x},\tau)\left[\hbar\frac{\partial}{\partial\tau}-\frac{\hbar^{2}}{2M}{\bf \Delta+V\left({\bf x}\right)}-\mu\right]\psi_{\alpha}({\bf x},\tau)\right.\right.\nonumber \\
 &  & \left.+\frac{1}{2}\int d\mathbf{x}'\psi_{\alpha}^{\ast}\left(\mathbf{x},\tau\right)\psi_{\alpha}\left(\mathbf{x},\tau\right)V^{({\rm int})}({\bf x-x'})\psi_{\alpha}^{\ast}\left(\mathbf{x'},\tau\right)\psi_{\alpha}\left(\mathbf{x'},\tau\right)\right\} \Bigg\}\nonumber \\
 &  & \times\overline{\exp\left\{ \int d\mathbf{x}\,\frac{-1}{\hbar}\int_{0}^{\hbar\beta}d\tau\sum_{\alpha=1}^{\mathcal{N}}\psi_{\alpha}^{*}({\bf x},\tau)\psi_{\alpha}({\bf x},\tau)U({\bf x})\right\} }\,,\label{ZP26}
\end{eqnarray}
where $\psi_{\alpha}^{*}({\bf x},\tau)$, $\psi_{\alpha}({\bf x},\tau)$
are the replica fields with the replica index $\alpha$. The remaining
disorder ensemble average of the exponential function can be performed
exactly on a formal level explained in Appendix A. Indeed, comparing
expressions (\ref{ZP26}) and (\ref{ZP10}) shows that averaging with
respect to the disorder potential $U({\bf x})$ corresponds to the
generating functional (\ref{ZP12}) with the auxiliary current field:
\begin{eqnarray}
j({\bf x})=\frac{-1}{\hbar}\int_{0}^{\hbar\beta}d\tau\sum_{\alpha=1}^{\mathcal{N}}\psi_{\alpha}^{*}({\bf x},\tau)\psi_{\alpha}({\bf x},\tau)\,.\label{ZP27}
\end{eqnarray}
Therefore, the disordered Bose gas is described by the disorder averaged,
replicated grand-canonical partition function 
\begin{eqnarray}
\overline{{\cal Z}^{\mathcal{N}}}=\left\{ \prod_{\alpha=1}^{\mathcal{N}}\oint{\cal D}\psi_{\alpha}^{*}\oint{\cal D}\psi_{\alpha}\right\} \, e^{-{\cal A}^{(\mathcal{N})}[\psi^{*},\psi]/\hbar},\label{ZP24}
\end{eqnarray}
with the following replica action 
\begin{eqnarray}
{\cal A}^{\left(\mathcal{N}\right)}[\psi^{*},\psi] & = & \int_{0}^{\hbar\beta}d\tau\int d\mathbf{x}\sum_{\alpha=1}^{\mathcal{N}}\,\left\{ \psi_{\alpha}^{*}({\bf x},\tau)\left[\hbar\frac{\partial}{\partial\tau}-\frac{\hbar^{2}}{2M}{\bf \Delta}+V({\bf x})-\mu\right]\psi_{\alpha}({\bf x},\tau)\right.\nonumber \\
 &  & \left.+\frac{1}{2}\int d\mathbf{x}'\psi_{\alpha}^{\ast}\left(\mathbf{x},\tau\right)\psi_{\alpha}\left(\mathbf{x},\tau\right)V^{({\rm int})}({\bf x-x'})\psi_{\alpha}^{\ast}\left(\mathbf{x'},\tau\right)\psi_{\alpha}\left(\mathbf{x'},\tau\right)\right\} \nonumber \\
 &  & +\sum_{i=2}^{\infty}\frac{1}{i!}\left(\frac{-1}{\hbar}\right)^{i-1}\int_{0}^{\hbar\beta}d\tau_{1}\cdots\int_{0}^{\hbar\beta}d\tau_{i}\int d\mathbf{x}_{1}\cdots\int d\mathbf{x}_{i}\nonumber \\
 &  & \times\sum_{\alpha_{1}=1}^{\mathcal{N}}\cdots\sum_{\alpha_{i}=1}^{\mathcal{N}}D^{\left(i\right)}(\mathbf{x}_{1},\ldots,\mathbf{x}_{i})\left|\psi_{\alpha_{1}}\left(\mathbf{x}_{1},\tau_{1}\right)\right|^{2}\cdots\left|\psi_{\alpha_{i}}\left(\mathbf{x}_{i},\tau_{i}\right)\right|^{2},\label{ZP29}
\end{eqnarray}
where $D^{\left(i\right)}(\mathbf{x}_{1},\ldots,\mathbf{x}_{i})$
denote the respective cumulants of the disorder potential, see Appendix
A. For any experimental realistic disorder potential the dominant
cumulant is of second order, as we assume, without loss of generality,
that the first cumulant vanishes according to \eqref{ZP1}. Therefore,
it is physically justified to restrict ourselves in the following
to the second cumulant, i.e., only $D^{\left(2\right)}(\mathbf{x}_{1}-\mathbf{x}_{2})=D(\mathbf{x}_{1}-\mathbf{x}_{2})$
contributes to the replicated action \eqref{ZP29}: 
\begin{eqnarray}
{\cal A}^{\left(\mathcal{N}\right)}[\psi^{*},\psi] & = & \int_{0}^{\hbar\beta}d\tau\int d\mathbf{x}\sum_{\alpha=1}^{\mathcal{N}}\,\left\{ \psi_{\alpha}^{*}({\bf x},\tau)\left[\hbar\frac{\partial}{\partial\tau}-\frac{\hbar^{2}}{2M}{\bf \Delta}+V({\bf x})-\mu\right]\psi_{\alpha}({\bf x},\tau)\right.\nonumber \\
 &  & \left.+\frac{1}{2}\int d\mathbf{x}'\psi_{\alpha}^{\ast}\left(\mathbf{x},\tau\right)\psi_{\alpha}\left(\mathbf{x},\tau\right)V^{({\rm int})}({\bf x-x'})\psi_{\alpha}^{\ast}\left(\mathbf{x'},\tau\right)\psi_{\alpha}\left(\mathbf{x'},\tau\right)\right\} \nonumber \\
 &  & -\frac{1}{2\hbar}\int_{0}^{\hbar\beta}d\tau\int_{0}^{\hbar\beta}d\tau'\int d\mathbf{x}\int d\mathbf{x}'\sum_{\alpha=1}^{\mathcal{N}}\sum_{\alpha'=1}^{\mathcal{N}}D({\bf x}-{\bf x'})\psi_{\alpha}^{*}({\bf x},\tau)\psi_{\alpha}({\bf x},\tau)\psi_{\alpha'}^{*}({\bf x'},\tau')\psi_{\alpha'}({\bf x'},\tau')\,.\label{28}
\end{eqnarray}
Thus, we conclude that, in this case, disorder leads to a residual
attractive interaction between the replica fields $\psi_{\alpha}^{*}({\bf x},\tau)$,
$\psi_{\alpha}({\bf x},\tau)$ which is, in general, bilocal in both
space and imaginary time.

\section{Hartree-Fock Mean-Field Equations}

Now we apply standard methods for developing a self-consistent mean-field
approximation \cite{Intro-95,Intro-96} in order to derive Hartree-Fock
mean-field equations for the Bose gas in a random potential. To this
end we use the Bogoliubov approximation, i.e., we split the Bose fields
$\psi_{\alpha}^{*}({\bf x},\tau)$, $\psi_{\alpha}({\bf x},\tau)$
into the background fields $\Psi_{\alpha}^{*}({\bf x},\tau)$, $\Psi_{\alpha}({\bf x},\tau)$
describing the condensate wave function, plus the fluctuations $\delta\psi_{\alpha}^{*}({\bf x},\tau)$,
$\delta\psi_{\alpha}({\bf x},\tau)$ describing the non-condensed
fractions: 
\begin{eqnarray}
\psi_{\alpha}^{*}({\bf x},\tau)=\Psi_{\alpha}^{*}({\bf x},\tau)+\delta\psi_{\alpha}^{*}({\bf x},\tau)\,,\hspace*{0.5cm}\psi_{\alpha}({\bf x},\tau)=\Psi_{\alpha}({\bf x},\tau)+\delta\psi_{\alpha}({\bf x},\tau)\,.\label{26}
\end{eqnarray}
Thus, the replica action (\ref{28}) decomposes according to ${\cal A}^{(\mathcal{N})}[\psi^{*},\psi]=\sum_{k=0}^{4}{\cal A}^{(\mathcal{N},k)}[\delta\psi^{*},\delta\psi]\,,$
where ${\cal A}^{(\mathcal{N},k)}[\delta\psi^{*},\delta\psi]$ denotes
all terms that contain fluctuations $\delta\psi_{\alpha}^{*}({\bf x},\tau)$,
$\delta\psi_{\alpha}({\bf x},\tau)$ to the $k^{{\rm {th}}}$ power.
Then, we approximate the higher nonlinear terms $k=3$ and $k=4$
within a Gaussian factorization, where expectation values are calculated
with respect to a fluctuation action $\tilde{{\cal A}}^{(\mathcal{N},2)}[\delta\psi^{*},\delta\psi]$
which is determined self-consistently below: 
\begin{eqnarray}
\langle\hspace*{0.3cm}\bullet\hspace*{3mm}\rangle=\frac{{\displaystyle \left\{ \prod_{\alpha=1}^{\mathcal{N}}\oint{\cal D}\delta\psi_{\alpha}^{*}\oint{\cal D}\delta\psi_{\alpha}\right\} \hspace*{3mm}\bullet\hspace*{3mm}e^{-\tilde{{\cal A}}^{(\mathcal{N},2)}[\delta\psi^{*},\delta\psi]/\hbar}}}{{\displaystyle \left\{ \prod_{\alpha=1}^{\mathcal{N}}\oint{\cal D}\delta\psi_{\alpha}^{*}\oint{\cal D}\delta\psi_{\alpha}\right\} e^{-\tilde{{\cal A}}^{(\mathcal{N},2)}[\delta\psi^{*},\delta\psi]/\hbar}}}\,.
\end{eqnarray}
As we restrict ourselves to a Hartree-Fock mean-field theory, we only
keep normal correlations $\langle\delta\psi_{\alpha}({\bf x},\tau)\,\delta\psi_{\alpha'}^{*}({\bf x'},\tau')\rangle$
and neglect all anomalous correlations of the form $\langle\delta\psi_{\alpha}({\bf x},\tau)\,\delta\psi_{\alpha'}({\bf x'},\tau')\rangle$
or $\langle\delta\psi_{\alpha}^{*}({\bf x},\tau)\,\delta\psi_{\alpha'}^{*}({\bf x'},\tau')\rangle$.
With this we obtain for the cubic terms in the fluctuations: 
\begin{eqnarray}
\delta\psi_{\alpha}^{*}({\bf x},\tau)\,\delta\psi_{\alpha}({\bf x},\tau)\,\delta\psi_{\alpha'}({\bf x'},\tau')\approx\langle\delta\psi_{\alpha}^{*}({\bf x},\tau^{+})\,\delta\psi_{\alpha}({\bf x},\tau)\rangle\,\delta\psi_{\alpha'}({\bf x'},\tau')+\langle\delta\psi_{\alpha}^{*}({\bf x},\tau)\,\delta\psi_{\alpha'}({\bf x'},\tau')\rangle\,\delta\psi_{\alpha}({\bf x},\tau),\label{29}
\end{eqnarray}
together with its complex conjugate and, correspondingly, the fourth
order terms in the fluctuations reduce to: 
\begin{eqnarray}
 &  & \delta\psi_{\alpha}^{*}({\bf x},\tau)\,\delta\psi_{\alpha}({\bf x},\tau)\,\delta\psi_{\alpha'}^{*}({\bf x'},\tau')\,\delta\psi_{\alpha'}({\bf x'},\tau')\label{31}\\
 &  & \approx\langle\delta\psi_{\alpha}^{*}({\bf x},\tau^{+})\,\delta\psi_{\alpha}({\bf x},\tau)\rangle\,\delta\psi_{\alpha'}^{*}({\bf x'},\tau')\,\delta\psi_{\alpha'}({\bf x'},\tau')+\langle\delta\psi_{\alpha'}^{*}({\bf x'},\tau'{}^{+})\,\delta\psi_{\alpha'}({\bf x'},\tau')\rangle\,\delta\psi_{\alpha}^{*}({\bf x},\tau)\,\delta\psi_{\alpha}({\bf x},\tau)\nonumber \\
 &  & +\langle\delta\psi_{\alpha}^{*}({\bf x},\tau)\,\delta\psi_{\alpha'}({\bf x'},\tau')\rangle\,\delta\psi_{\alpha}({\bf x},\tau)\,\delta\psi_{\alpha'}^{*}({\bf x'},\tau')+\langle\delta\psi_{\alpha}({\bf x},\tau)\,\delta\psi_{\alpha'}^{*}({\bf x'},\tau'))\rangle\,\delta\psi_{\alpha}^{*}({\bf x},\tau)\,\delta\psi_{\alpha'}({\bf x'},\tau')\nonumber \\
 &  & -\langle\delta\psi_{\alpha}^{*}({\bf x},\tau^{+})\,\delta\psi_{\alpha}({\bf x},\tau)\rangle\,\langle\delta\psi_{\alpha'}^{*}({\bf x'},\tau'^{+})\,\delta\psi_{\alpha'}({\bf x'},\tau')\rangle-\langle\delta\psi_{\alpha}^{*}({\bf x},\tau)\,\delta\psi_{\alpha'}({\bf x'},\tau')\rangle\,\langle\delta\psi_{\alpha}({\bf x},\tau)\,\delta\psi_{\alpha'}^{*}({\bf x'},\tau')\rangle\,.\nonumber 
\end{eqnarray}

Here we have used $\tau^{+}$ as an imaginary time which is infinitesimally
later than $\tau$ in order to guarantee the normal ordering of the
fluctuations within the respective expectation values. Therefore,
the Gaussian factorization procedure for a Hartree-Fock mean-field
theory leads to the following approximation of the replica action
\eqref{28}: 
\begin{eqnarray}
{\cal A}^{(\mathcal{N})}[\psi^{*},\psi]\approx\tilde{{\cal A}}^{(\mathcal{N},0)}[\delta\psi^{*},\delta\psi]+\tilde{{\cal A}}^{(\mathcal{N},1)}[\delta\psi^{*},\delta\psi]\,+\tilde{{\cal A}}^{(\mathcal{N},2)}[\delta\psi^{*},\delta\psi]\,,\label{ZP30}
\end{eqnarray}
where $\tilde{{\cal A}}^{(\mathcal{N},k)}[\delta\psi^{*},\delta\psi]$
denotes the $k^{{\rm {th}}}$-order terms of the replica action \eqref{28}.
To make our notation concise, we express in all those terms the fluctuations
in \eqref{29}, \eqref{31} by the following mean-fields: 
\begin{eqnarray}
Q_{\alpha\alpha'}({\bf x},\tau;{\bf x'},\tau') & = & \Psi_{\alpha}({\bf x},\tau)\,\Psi_{\alpha'}^{*}({\bf x'},\tau')+\langle\delta\psi_{\alpha}({\bf x},\tau)\,\delta\psi_{\alpha'}^{*}({\bf x'},\tau')\rangle\,,\label{MMM1}\\
Q_{\alpha\alpha'}^{*}({\bf x},\tau;{\bf x'},\tau') & = & Q_{\alpha'\alpha}({\bf x'},\tau';{\bf x},\tau)\,,\label{MMM2}\\
\Sigma_{\alpha}({\bf x},\tau) & = & Q_{\alpha\alpha}({\bf x},\tau;{\bf x},\tau^{+})\,.\label{MMM3}
\end{eqnarray}
With this the first term of the replica action (\ref{ZP30}), which
is independent of the fluctuations $\delta\psi_{\alpha}^{*}({\bf x},\tau)$,
$\delta\psi_{\alpha}({\bf x},\tau)$, reads:

\begin{alignat}{1}
\tilde{{\cal A}}^{(\mathcal{N},0)}[\delta\psi^{*},\delta\psi]= & \int_{0}^{\hbar\beta}d\tau\int d\mathbf{x}\sum_{\alpha=1}^{\mathcal{N}}\,\Bigg\{\Psi_{\alpha}^{*}({\bf x},\tau)\left[\hbar\frac{\partial}{\partial\tau}-\frac{\hbar^{2}}{2M}{\bf \Delta}+V({\bf x})-\mu\right]\Psi_{\alpha}({\bf x},\tau)\nonumber \\
 & -\frac{1}{2}\int d\mathbf{x}'V^{({\rm int})}({\bf x-x'})\Biggl[\Psi_{\alpha}^{*}({\bf x},\tau)\Psi_{\alpha}({\bf x},\tau)\Psi_{\alpha}^{*}({\bf x'},\tau)\Psi_{\alpha}({\bf x'},\tau)+\Sigma_{\alpha}({\bf x},\tau)\,\Sigma_{\alpha}({\bf x'},\tau)\nonumber \\
 & -2\Sigma_{\alpha}({\bf x},\tau)\,\Psi_{\alpha}^{*}({\bf x'},\tau)\Psi_{\alpha}({\bf x'},\tau)+Q_{\alpha\alpha}({\bf x},\tau;{\bf x'},\tau)Q_{\alpha\alpha}^{*}({\bf x},\tau;{\bf x'},\tau)\nonumber \\
 & -Q_{\alpha\alpha}({\bf x},\tau;{\bf x'},\tau)\Psi_{\alpha}({\bf x'},\tau)\Psi_{\alpha}^{*}({\bf x},\tau)-Q_{\alpha\alpha}^{*}({\bf x},\tau;{\bf x'},\tau)\Psi_{\alpha}({\bf x},\tau)\Psi_{\alpha}^{*}({\bf x'},\tau)\Biggr]\Biggr\}\nonumber \\
 & +\frac{1}{2\hbar}\,\int_{0}^{\hbar\beta}d\tau\int_{0}^{\hbar\beta}d\tau'\int d\mathbf{x}\int d\mathbf{x}'\sum_{\alpha=1}^{\mathcal{N}}\sum_{\alpha'=1}^{\mathcal{N}}\, D({\bf x}-{\bf x'})\nonumber \\
 & \times\Bigg\{\Psi_{\alpha}^{*}({\bf x},\tau)\Psi_{\alpha}({\bf x},\tau)\Psi_{\alpha'}^{*}({\bf x'},\tau')\Psi_{\alpha'}({\bf x'},\tau')+\Sigma_{\alpha}({\bf x},\tau)\,\Sigma_{\alpha'}({\bf x'},\tau')\nonumber \\
 & -2\Sigma_{\alpha}({\bf x},\tau)\,\Psi_{\alpha'}^{*}({\bf x'},\tau')\Psi_{\alpha'}({\bf x'},\tau')+Q_{\alpha\alpha'}({\bf x},\tau;{\bf x'},\tau')Q_{\alpha\alpha'}^{*}({\bf x},\tau;{\bf x'},\tau')\nonumber \\
 & -Q_{\alpha\alpha'}({\bf x},\tau;{\bf x'},\tau')\Psi_{\alpha'}({\bf x'},\tau')\Psi_{\alpha}^{*}({\bf x},\tau)-Q_{\alpha\alpha'}^{*}({\bf x},\tau;{\bf x'},\tau')\Psi_{\alpha}({\bf x},\tau)\Psi_{\alpha'}^{*}({\bf x'},\tau')\Bigg\}\,.\label{ZERO}
\end{alignat}
Furthermore, the second term of decomposition (\ref{ZP30}), i.e.,
$\tilde{{\cal A}}^{(\mathcal{N},1)}[\delta\psi^{*},\delta\psi]$,
is linear in the fluctuations $\delta\psi_{\alpha}^{*}({\bf x},\tau)$,
$\delta\psi_{\alpha}({\bf x},\tau)$ and turns out to vanish. Indeed,
following the field-theoretic background field method \cite{Axel-29,Axel-30}
it can be shown that the first-order terms $\tilde{{\cal A}}^{(\mathcal{N},1)}[\delta\psi^{*},\delta\psi]$
can be neglected here as they would vanish later on from extremising
$\tilde{{\cal A}}^{(\mathcal{N},0)}[\delta\psi^{*},\delta\psi]$ with
respect to the background fields $\Psi_{\alpha}^{*}({\bf x},\tau)$,
$\Psi_{\alpha}({\bf x},\tau)$. The third term of decomposition (\ref{ZP30})
is quadratic in the fluctuations: 
\begin{eqnarray}
\tilde{{\cal A}}^{(\mathcal{N},2)}[\delta\psi^{*},\delta\psi] & = & \int_{0}^{\hbar\beta}d\tau\int d\mathbf{x}\sum_{\alpha=1}^{\mathcal{N}}\,\Bigg\{\delta\psi_{\alpha}^{*}({\bf x},\tau)\left[\hbar\frac{\partial}{\partial\tau}-\frac{\hbar^{2}}{2M}{\bf \Delta}+V({\bf x})-\mu\right]\delta\psi_{\alpha}({\bf x},\tau)\nonumber \\
 &  & +\frac{1}{2}\int d\mathbf{x}'V^{({\rm int})}({\bf x-x'})\Biggl[2\Sigma_{\alpha}({\bf x},\tau)\,\delta\psi_{\alpha}^{*}({\bf x'},\tau)\,\delta\psi_{\alpha}({\bf x'},\tau)\nonumber \\
 &  & +Q_{\alpha\alpha}({\bf x},\tau;{\bf x'},\tau)\,\delta\psi_{\alpha}({\bf x'},\tau)\,\delta\psi_{\alpha}^{*}({\bf x},\tau)+Q_{\alpha\alpha}^{*}({\bf x},\tau;{\bf x'},\tau)\,\delta\psi_{\alpha}({\bf x},\tau)\,\delta\psi_{\alpha}^{*}({\bf x'},\tau)\Biggr]\Bigg\}\nonumber \\
 &  & -\frac{1}{2\hbar}\,\int_{0}^{\hbar\beta}d\tau\int_{0}^{\hbar\beta}d\tau'\int d\mathbf{x}\int d\mathbf{x}'\sum_{\alpha=1}^{\mathcal{N}}\sum_{\alpha'=1}^{\mathcal{N}}\, D({\bf x}-{\bf x'})\nonumber \\
 &  & \times\Bigg\{2\,\Sigma_{\alpha}({\bf x},\tau)\,\delta\psi_{\alpha'}^{*}({\bf x'},\tau')\,\delta\psi_{\alpha'}({\bf x'},\tau')+Q_{\alpha\alpha'}({\bf x},\tau;{\bf x'},\tau')\,\delta\psi_{\alpha'}({\bf x'},\tau')\,\delta\psi_{\alpha}^{*}({\bf x},\tau)\nonumber \\
 &  & +Q_{\alpha\alpha'}^{*}({\bf x},\tau;{\bf x'},\tau')\,\delta\psi_{\alpha}({\bf x},\tau)\,\delta\psi_{\alpha'}^{*}({\bf x'},\tau')\Bigg\}\,.\label{FLUA}
\end{eqnarray}

Inserting expression (\ref{ZP30}) together with above results (\ref{ZERO})
and (\ref{FLUA}), into formula (\ref{ZP24}) leads to the replicated
effective potential: 
\begin{eqnarray}
V_{{\rm eff}}^{(\mathcal{N})}=-\frac{1}{\beta}\ln\overline{{\cal Z}^{\mathcal{N}}},\label{37}
\end{eqnarray}
which is given by: 
\begin{eqnarray}
V_{{\rm eff}}^{(\mathcal{N})}=\frac{\tilde{{\cal A}}^{(\mathcal{N},0)}[\delta\psi^{*},\delta\psi]}{\hbar\beta}-\frac{1}{\beta}\ln\left\{ \left[\prod_{\alpha=1}^{\mathcal{N}}\oint{\cal D}\delta\psi_{\alpha}^{*}\oint{\cal D}\delta\psi_{\alpha}\right]e^{-\tilde{{\cal A}}^{(\mathcal{N},2)}[\delta\psi^{*},\delta\psi]/\hbar}\right\} \,.\label{EFF1}
\end{eqnarray}
It represents a functional of all mean-fields: $V_{{\rm eff}}^{(\mathcal{N})}=V_{{\rm eff}}^{(\mathcal{N})}[\Psi^{*},\,\Psi,\, Q^{*},\, Q,\,\Sigma]$.
Extremising expression (\ref{EFF1}) with respect to the mean-fields
$Q_{\alpha\alpha'}^{*}({\bf x},\tau;{\bf x'},\tau')$, $Q_{\alpha\alpha'}({\bf x},\tau;{\bf x'},\tau')$,
and $\Sigma_{\alpha}({\bf x},\tau)$ reproduces their definitions
(\ref{MMM1})--(\ref{MMM3}), where the expectation values turn out
to be calculated with respect to the fluctuation action (\ref{FLUA}).
Furthermore, an extremisation of the replicated effective potential
(\ref{EFF1}) with respect to the background fields $\Psi_{\alpha}^{*}({\bf x},\tau)$,
$\Psi_{\alpha}({\bf x},\tau)$ leads to the Gross-Pitaevskii equation:

\begin{eqnarray}
\negthickspace\negthickspace\negthickspace\negthickspace\negthickspace\negthickspace\negthickspace\negthickspace &  & \left\{ \hbar\frac{\partial}{\partial\tau}-\frac{\hbar^{2}}{2M}{\bf \Delta}+V({\bf x})-\mu\right\} \Psi_{\alpha}({\bf x},\tau)-\int d\mathbf{x}'V^{({\rm int})}({\bf x-x'})\nonumber \\
 &  & \times\Biggl[\Psi_{\alpha}({\bf x},\tau)\Psi_{\alpha}^{*}({\bf x'},\tau)\Psi_{\alpha}({\bf x'},\tau)-\Sigma_{\alpha}({\bf x},\tau)\,\Psi_{\alpha}({\bf x'},\tau)-Q_{\alpha\alpha}({\bf x},\tau;{\bf x'},\tau)\Psi_{\alpha}({\bf x'},\tau)\Biggr]\label{MF1}\\
\negthickspace\negthickspace\negthickspace\negthickspace\negthickspace\negthickspace\negthickspace\negthickspace &  & =\frac{1}{\hbar}\int_{0}^{\hbar\beta}d\tau'\int d\mathbf{x}'\sum_{\alpha'=1}^{N}D({\bf x}-{\bf x'})\Biggl\{ Q_{\alpha\alpha'}({\bf x},\tau;{\bf x'},\tau')\Psi_{\alpha'}({\bf x'},\tau')+\Biggl[\Sigma_{\alpha'}({\bf x'},\tau')-\Psi_{\alpha'}({\bf x'},\tau')\Psi_{\alpha'}^{*}({\bf x'},\tau')\Biggr]\Psi_{\alpha}({\bf x},\tau)\Biggr\}\nonumber 
\end{eqnarray}
and its complex conjugate.

\section{Replica Symmetry}

Now we apply the replica symmetry, where we assume that all the respective
replica indices $\alpha$ contribute in the same way. Furthermore,
the dirty boson problem is translationally invariant in imaginary
time. With this we get for the background 
\begin{equation}
\Psi_{\alpha}({\bf x},\tau)=\Psi({\bf x}),\;\;\;\;\Psi_{\alpha}^{*}({\bf x},\tau)=\Psi^{*}({\bf x}),\;\;\;\;\Sigma_{\alpha}({\bf x},\tau)=\Sigma({\bf x}),\label{RS}
\end{equation}
and for the mean fields 
\begin{eqnarray}
Q_{\alpha\alpha'}({\bf x},\tau;{\bf x'},\tau')=Q\left({\bf x}-{\bf x'},\frac{{\bf x}+{\bf x'}}{2};\tau-\tau'\right)\,\delta_{\alpha\alpha'}+q\left({\bf x}-{\bf x'},\frac{{\bf x}+{\bf x'}}{2};\tau-\tau'\right)+\Psi^{*}({\bf x})\Psi({\bf x})\,,\label{AN1}
\end{eqnarray}
and its complex conjugate. In \eqref{AN1} we perform a Fourier-Matsubara
decomposition with respect to the differences in space and time, i.e.,
${\bf x}-{\bf x'}$ and $\tau-\tau'.$ Furthermore, we assume within
a semi-classical approximation that the dependence on the center of
mass coordinate $\left({\bf x}+{\bf x'}\right)/2$ is smooth, so we
get

\begin{equation}
Q\left({\bf x}-{\bf x'},\frac{{\bf x}+{\bf x'}}{2};\tau-\tau'\right)=\int\frac{d{\bf k}}{(2\pi)^{n}}\, e^{i{\bf k}({\bf x}-{\bf x'})}\frac{1}{\hbar\beta}\sum_{m=-\infty}^{\infty}e^{-i\omega_{m}(\tau-\tau')}Q_{m}\left({\bf k},\frac{{\bf x}+{\bf x'}}{2}\right),\label{AN2}
\end{equation}
\begin{equation}
q\left({\bf x}-{\bf x'},\frac{{\bf x}+{\bf x'}}{2};\tau-\tau'\right)=\int\frac{d{\bf k}}{(2\pi)^{n}}\, e^{i{\bf k}({\bf x}-{\bf x'})}\frac{1}{\hbar\beta}\sum_{m=-\infty}^{\infty}e^{-i\omega_{m}(\tau-\tau')}q_{m}\left({\bf k},\frac{{\bf x}+{\bf x'}}{2}\right),\label{AN2-1}
\end{equation}
and their complex conjugates, where $\omega_{m}=2\pi m/\hbar\beta$
denote the bosonic Matsubara frequencies and $\mathbf{k}$ the wave
vector.

Using this ansatz, we have to evaluate the expectation values in the
mean-field equations (\ref{MMM1})--(\ref{MMM3}) and (\ref{MF1}).
To this end we note that the fluctuation action (\ref{FLUA}) is of
the general form 
\begin{alignat}{1}
\tilde{{\cal A}}^{(\mathcal{N},2)}[\delta\psi^{*},\delta\psi]= & \int_{0}^{\hbar\beta}\hspace*{-1mm}d\tau\int_{0}^{\hbar\beta}\hspace*{-1mm}d\tau'\int d{\bf x}\int d{\bf x}'\sum_{\alpha=1}^{\mathcal{N}}\sum_{\alpha'=1}^{\mathcal{N}}\nonumber \\
 & \frac{1}{2}\Big(\delta\psi_{\alpha}^{*}({\bf x},\tau),\,\delta\psi_{\alpha}({\bf x},\tau)\Big)G_{\alpha\alpha'}^{-1}\left({\bf x}-{\bf x'},\frac{{\bf x}+{\bf x'}}{2};\tau-\tau'\right)\left(\begin{array}{@{}c}
\delta\psi_{\alpha'}({\bf x'},\tau')\\
\delta\psi_{\alpha'}^{*}({\bf x'},\tau')
\end{array}\right),\label{2.44}
\end{alignat}
where the semi-classical Fourier-Matsubara transformation of the integral
kernel 
\begin{eqnarray}
G_{\alpha\alpha'}^{-1}\left({\bf x}-{\bf x'},\frac{{\bf x}+{\bf x'}}{2};\tau-\tau'\right)=\int\frac{d{\bf k}}{(2\pi)^{n}}\, e^{i{\bf k}({\bf x}-{\bf x'})}\frac{1}{\hbar\beta}\sum_{m=-\infty}^{\infty}e^{-i\omega_{m}(\tau-\tau')}\, G_{\alpha\alpha'}^{-1}\left({\bf k},\omega_{m},\frac{{\bf x}+{\bf x'}}{2}\right)\,,\label{FMT}
\end{eqnarray}
decomposes according to 
\begin{alignat}{1}
G_{\alpha\alpha'}^{-1}\left({\bf k},\omega_{m},\frac{{\bf x}+{\bf x'}}{2}\right)=\left(\begin{array}{@{}cc}
a({\bf k},\omega_{m},\frac{{\bf x}+{\bf x'}}{2}) & 0\\
0 & a^{*}({\bf k},\omega_{m},\frac{{\bf x}+{\bf x'}}{2})
\end{array}\right)\,\delta_{\alpha\alpha'}+\left(\begin{array}{@{}cc}
b({\bf k},\omega_{m},\frac{{\bf x}+{\bf x'}}{2}) & 0\\
0 & b^{*}({\bf k},\omega_{m},\frac{{\bf x}+{\bf x'}}{2})
\end{array}\right),\nonumber \\
\label{DEKO4}
\end{alignat}
with the abbreviations 
\begin{alignat}{1}
a\left({\bf k},\omega_{m},\frac{{\bf x}+{\bf x'}}{2}\right)= & -i\hbar\omega_{m}+\epsilon({\bf k})-\mu+\Sigma\left(\frac{{\bf x}+{\bf x'}}{2}\right)\int d^{n}x''V^{({\rm int})}({\bf x''})+V^{({\rm int})}({\bf k})\Psi^{*}\left(\frac{{\bf x}+{\bf x'}}{2}\right)\Psi\left(\frac{{\bf x}+{\bf x'}}{2}\right)\nonumber \\
 & +V\left(\frac{{\bf x}+{\bf x'}}{2}\right)-\frac{1}{\hbar}\int\frac{d{\bf k}'}{\left(2\pi\right)^{n}}D\left({\bf k'}\right)Q_{m}\left({\bf k}-{\bf k'},\frac{{\bf x}+{\bf x'}}{2}\right)-\mathcal{N}\beta D\left({\bf k}\right)\Sigma\left(\frac{{\bf x}+{\bf x'}}{2}\right)\,\delta_{m,0}\nonumber \\
 & +\int\frac{d{\bf k}'}{\left(2\pi\right)^{n}}V^{({\rm int})}({\bf k'})\left[q_{m}\left({\bf k}-{\bf k'},\frac{{\bf x}+{\bf x'}}{2}\right)+Q_{m}\left({\bf k}-{\bf k'},\frac{{\bf x}+{\bf x'}}{2}\right)\right]\,,\label{AB}
\end{alignat}
\begin{alignat}{1}
b\left({\bf k},\omega_{m},\frac{{\bf x}+{\bf x'}}{2}\right)=-\frac{1}{\hbar}\left[\int\frac{d{\bf k}'}{\left(2\pi\right)^{n}}D\left({\bf k'}\right)q_{m}\left({\bf k}-{\bf k'},\frac{{\bf x}+{\bf x'}}{2}\right)+\hbar\beta D\left({\bf k}\right)\Psi^{*}\left(\frac{{\bf x}+{\bf x'}}{2}\right)\Psi\left(\frac{{\bf x}+{\bf x'}}{2}\right)\,\delta_{m,0}\right],\nonumber \\
\label{BA}
\end{alignat}
and the free dispersion $\epsilon({\bf k})=\hbar^{2}{\bf k}^{2}/2M\,.$
Furthermore, $D\left({\bf k}\right)$ and $V^{({\rm int})}({\bf k})$
are the Fourier transforms of the disorder correlation function $D({\bf x})$
and the two-particle interaction potential $V^{({\rm int})}({\bf x})$,
respectively: $D({\bf x})=\int\frac{d{\bf k}}{(2\pi)^{n}}\, D({\bf k})e^{i{\bf k}{\bf x}},$
$V^{({\rm int})}({\bf \mathbf{x}})=\int\frac{d{\bf k}}{(2\pi)^{n}}\, V^{({\rm int})}({\bf k})e^{i{\bf k}{\bf x}}.$

The corresponding Green function follows from solving 
\begin{alignat}{1}
 & \int_{0}^{\hbar\beta}d\tau\int d{\bf x}\sum_{\alpha=1}^{\mathcal{N}}G_{\alpha_{1}\alpha}^{-1}\left({\bf x}-{\bf x}_{1},\frac{{\bf x}+{\bf x}_{1}}{2};\tau-\tau_{1}\right)\, G_{\alpha\alpha_{2}}\left({\bf x}_{2}-{\bf x},\frac{{\bf x}+{\bf x}_{2}}{2};\tau_{2}-\tau\right)\nonumber \\
 & =\hbar\delta({\bf x}_{1}-{\bf x}_{2})\delta(\tau_{1}-\tau_{2})\delta_{\alpha_{1}\alpha_{2}},
\end{alignat}
which reduces with a semi-classical Fourier-Matsubara transformation
to the algebraic identity: 
\begin{eqnarray}
\sum_{\alpha=1}^{\mathcal{N}}G_{\alpha_{1}\alpha}^{-1}\left({\bf k},\omega_{m},\frac{{\bf x}+{\bf x'}}{2}\right)\, G_{\alpha\alpha_{2}}\left({\bf k},\omega_{m},\frac{{\bf x}+{\bf x'}}{2}\right)=\hbar\,\delta_{\alpha_{1}\alpha_{2}}\,.
\end{eqnarray}
Thus, the corresponding Green function, which contains expectation
values according to 
\begin{eqnarray}
G_{\alpha\alpha'}\left({\bf x}-{\bf x'},\frac{{\bf x}+{\bf x'}}{2};\tau-\tau'\right)=\left(\begin{array}{@{}cc}
\langle\delta\psi_{\alpha}({\bf x},\tau)\delta\psi_{\alpha'}^{*}({\bf x'},\tau')\rangle & 0\\
0 & \langle\delta\psi_{\alpha}^{*}({\bf x},\tau)\delta\psi_{\alpha'}({\bf x'},\tau')\rangle
\end{array}\right)\,,
\end{eqnarray}
is determined from 
\begin{eqnarray}
\langle\delta\psi_{\alpha}({\bf x},\tau)\delta\psi_{\alpha'}^{*}({\bf x'},\tau')\rangle=g_{1}\left({\bf x}-{\bf x'},\frac{{\bf x}+{\bf x'}}{2};\tau-\tau'\right)\,\delta_{\alpha\alpha'}+g_{2}\left({\bf x}-{\bf x'},\frac{{\bf x}+{\bf x'}}{2};\tau-\tau'\right),\label{EXP}
\end{eqnarray}
with the contributions: 
\begin{eqnarray}
g_{1}\left({\bf x}-{\bf x'},\frac{{\bf x}+{\bf x'}}{2};\tau-\tau'\right) & = & \int\frac{d{\bf k}}{(2\pi)^{n}}\, e^{i{\bf k}({\bf x}-{\bf x'})}\sum_{m=-\infty}^{\infty}\frac{e^{-i\omega_{m}(\tau-\tau')}}{\beta a({\bf k},\omega_{m},\frac{{\bf x}+{\bf x'}}{2})}\,,\label{G1}\\
g_{2}\left({\bf x}-{\bf x'},\frac{{\bf x}+{\bf x'}}{2};\tau-\tau'\right) & = & \int\frac{d{\bf k}}{(2\pi)^{n}}\, e^{i{\bf k}({\bf x}-{\bf x'})}\,\sum_{m=-\infty}^{\infty}\frac{e^{-i\omega_{m}(\tau-\tau')}}{\beta\mathcal{N}}\nonumber \\
 &  & \times{\displaystyle \left[\frac{1}{\mathcal{N}b({\bf k},\omega_{m},\frac{{\bf x}+{\bf x'}}{2})+a({\bf k},\omega_{m},\frac{{\bf x}+{\bf x'}}{2})}-\frac{1}{a({\bf k},\omega_{m},\frac{{\bf x}+{\bf x'}}{2})}\right]}\,.\label{G2}
\end{eqnarray}
Comparing Eqs. (\ref{MMM1})--(\ref{MMM3}) and (\ref{AN1})--\eqref{AN2-1}
with \eqref{EXP}--\eqref{G2} yields: 
\begin{equation}
Q_{m}\left({\bf k},\frac{{\bf x}+{\bf x'}}{2}\right)=\frac{\hbar}{a({\bf k},\omega_{m},\frac{{\bf x}+{\bf x'}}{2})}\,,\label{74-1}
\end{equation}
\begin{equation}
q_{m}\left({\bf k},\frac{{\bf x}+{\bf x'}}{2}\right)=\frac{\hbar}{\mathcal{N}}{\displaystyle \left[\frac{1}{\mathcal{N}b({\bf k},\omega_{m},\frac{{\bf x}+{\bf x'}}{2})+a({\bf k},\omega_{m},\frac{{\bf x}+{\bf x'}}{2})}-\frac{1}{a({\bf k},\omega_{m},\frac{{\bf x}+{\bf x'}}{2})}\right]}\,,\label{75-1}
\end{equation}
and their complex conjugates. Equations \eqref{74-1} and \eqref{75-1}
represent, due to expressions \eqref{AB} and \eqref{BA}, two coupled
integral mean-field equations for the quantities $Q_{m}({\bf k},\frac{{\bf x}+{\bf x'}}{2})$
and $q_{m}({\bf k},\frac{{\bf x}+{\bf x'}}{2})$. As it is not possible
to solve them analytically for a general disorder potential and a
general interaction potential, we specialize now to a $\delta$-correlated
disorder potential and a contact interaction potential.

\section{Delta-correlated Disorder and Contact Interaction Potential}

Now we elaborate a solution of our mean-field equations for the special
case of a $\delta$-correlated disorder potential, which is defined
in Eq. (\ref{ZP4}), i.e., we have

\begin{equation}
D\left({\bf k}\right)=D,\label{Delta}
\end{equation}
where $D$ denotes the disorder strength. Furthermore, we choose a
contact interaction potential

\begin{equation}
V^{({\rm int})}({\bf x-x'})=g\delta(\mathbf{x}-\mathbf{x'}),\label{77}
\end{equation}
where $g$ denotes the interaction coupling strength. In this case
formulas \eqref{AB} and \eqref{BA} reduce to:

\begin{alignat}{1}
a\left({\bf k},\omega_{m},\frac{{\bf x}+{\bf x'}}{2}\right) & =-i\hbar\omega_{m}+\epsilon({\bf k})-\mu+2g\Sigma\left(\frac{{\bf x}+{\bf x'}}{2}\right)+V\left(\frac{{\bf x}+{\bf x'}}{2}\right)-\frac{D}{\hbar}Q_{m}\left(\frac{{\bf x}+{\bf x'}}{2}\right)-\mathcal{N}\beta D\Sigma\left(\frac{{\bf x}+{\bf x'}}{2}\right)\,\delta_{m,0},\label{74}
\end{alignat}
and 
\begin{alignat}{1}
b\left({\bf k},\omega_{m},\frac{{\bf x}+{\bf x'}}{2}\right)=-\frac{D}{\hbar}\left[q_{m}\left(\frac{{\bf x}+{\bf x'}}{2}\right)+\hbar\beta\Psi^{*}\left(\frac{{\bf x}+{\bf x'}}{2}\right)\Psi\left(\frac{{\bf x}+{\bf x'}}{2}\right)\,\delta_{m,0}\right],\label{AB-2}
\end{alignat}
where we have introduced the abbreviation 
\begin{equation}
Q_{m}\left(\frac{{\bf x}+{\bf x'}}{2}\right)=\int\frac{d{\bf k'}}{\left(2\pi\right)^{n}}Q_{m}\left({\bf k'},\frac{{\bf x}+{\bf x'}}{2}\right),\label{eq:*}
\end{equation}
\begin{equation}
q_{m}\left(\frac{{\bf x}+{\bf x'}}{2}\right)=\int\frac{d{\bf k'}}{\left(2\pi\right)^{n}}q_{m}\left({\bf k'},\frac{{\bf x}+{\bf x'}}{2}\right).\label{eq:**}
\end{equation}
Expressions \eqref{74-1} and \eqref{75-1} yield then together with
expressions \eqref{eq:*} and \eqref{eq:**} algebraic mean-field
equations, which we can solve. Inserting expressions \eqref{74} and
\eqref{AB-2} into Eqs. \eqref{74-1} and \eqref{75-1} and taking
${\bf x}={\bf x'}$ in expressions \eqref{eq:*} and \eqref{eq:**},
with the Schwinger integral \cite{Kleinert1} 
\begin{eqnarray}
\frac{1}{a^{\nu}}=\frac{1}{\Gamma(\nu)}\int_{0}^{\infty}ds\, s^{\nu-1}\, e^{-as},\label{SCH}
\end{eqnarray}
and formula \cite[(8.310.1)]{Gradshteyn}, we obtain the following
self-consistency equations:

\begin{alignat}{1}
Q_{m}({\bf x}) & =\Gamma\left(1-\frac{n}{2}\right)\hbar\left(\frac{M}{2\pi\hbar^{2}}\right)^{n/2}\left[-i\hbar\omega_{m}-\mu+2g\Sigma({\bf x})+V({\bf x})-\frac{D}{\hbar}Q_{m}({\bf x})-\mathcal{N}\beta D\Sigma({\bf x})\,\delta_{m,0}\right]^{\frac{n}{2}-1}\,,\nonumber \\
\label{Q-1}
\end{alignat}
\begin{alignat}{1}
q_{m}({\bf x})= & -\Gamma\left(1-\frac{n}{2}\right)\frac{\hbar}{\mathcal{N}}\left(\frac{M}{2\pi\hbar^{2}}\right)^{n/2}\left[-i\hbar\omega_{m}-\mu+2g\Sigma({\bf x})+V({\bf x})-\frac{D}{\hbar}Q_{m}({\bf x})-\mathcal{N}\beta D\Sigma({\bf x})\,\delta_{m,0}\right]^{\frac{n}{2}-1}\nonumber \\
 & +\Gamma\left(1-\frac{n}{2}\right)\frac{\hbar}{\mathcal{N}}\left(\frac{M}{2\pi\hbar^{2}}\right)^{n/2}\Biggl\{-i\hbar\omega_{m}-\mu+2g\Sigma({\bf x})+V({\bf x})-\frac{D}{\hbar}Q_{m}({\bf x})-\mathcal{N}\beta D\Sigma({\bf x})\,\delta_{m,0}\nonumber \\
 & -\mathcal{N}\frac{D}{\hbar}\left[q_{m}({\bf x})+\hbar\beta\Psi^{*}({\bf x})\Psi({\bf x})\,\delta_{m,0}\right]\Biggr\}^{\frac{n}{2}-1}\,.\label{qm-1}
\end{alignat}
From the above expressions, we conclude $Q_{m}^{*}({\bf x})=Q_{-m}({\bf x})$
and $q_{m}^{*}({\bf x})=q_{-m}({\bf x})\,.$ With this we read off
from Eqs. (\ref{AN2}) and (\ref{AN2-1}) that $Q^{*}({\bf x};\tau'-\tau)=Q({\bf x};\tau-\tau')$
and $q^{*}({\bf x};\tau'-\tau)=q({\bf x};\tau-\tau')$, respectively.

The expressions for $Q_{m}({\bf x})$ and $q_{m}({\bf x})$ in Eqs.
\eqref{Q-1} and \eqref{qm-1} turn out to diverge in two spatial
dimensions because of the prefactor $\Gamma\left(1-\frac{n}{2}\right)$.
This means that our theory in its actual form is not valid in the
two-dimensional case. In order to get valid self-consistency equations
also in two dimensions, one way would be to choose a disorder potential
with a finite correlation length, e.g., a Lorentzian-correlated potential.
Then this finite correlation length would provide a regularization
that would yield together with a renormalization, finite self-consistency
equations. As the treatment of a Lorentzian-correlated disorder potential
lies out of the scope of the present paper, we will restrict ourselves
later on to the study of the one- and the three-dimensional cases.

We note in Eqs. \eqref{Q-1} and \eqref{qm-1} that the terms containing
the parameter $\beta$ are always multiplied by the number of replicas
$\mathcal{N}$. This is important because it means that in the zero
temperature case, i.e., $\beta\rightarrow\infty$, those terms will
be eliminated in the replica limit $\mathcal{N}\to0$, and otherwise
they would diverge.

In Ref. \cite{Intro-90} the replica limit is taken as soon as the
replica number $\mathcal{N}$ appears at different steps of the calculation.
In our work, and contrary to Ref. \cite{Intro-90}, until this level
of the calculation no replica limit was performed. We are taking this
limit as late as possible in order to avoid any loss of terms due
to the performance of the replica limit in the earlier steps of the
calculation.

Note that in the replica limit $\mathcal{N}\to0$, Eqs. \eqref{Q-1}
and \eqref{qm-1} yield 
\begin{equation}
Q_{m}({\bf x})=\Gamma\left(1-\frac{n}{2}\right)\hbar\left(\frac{M}{2\pi\hbar^{2}}\right)^{n/2}\left[-i\hbar\omega_{m}-\mu+2g\Sigma({\bf x})+V({\bf x})-\frac{D}{\hbar}Q_{m}({\bf x})\right]^{\frac{n}{2}-1}\label{Qg}
\end{equation}
and 
\begin{equation}
q({\bf x})=D\Gamma\left(2-\frac{n}{2}\right)\,\left(\frac{M}{2\pi\hbar^{2}}\right)^{n/2}\frac{\left[q({\bf x})+\Psi^{*}({\bf x})\Psi({\bf x})\right]}{\left[-\mu+2g\Sigma({\bf x})+V({\bf x})-\frac{D}{\hbar}Q_{0}({\bf x})\right]^{2-\frac{n}{2}}},\label{qg}
\end{equation}
where $q({\bf x})=q_{0}({\bf x})/\hbar\beta$ and $q_{m}({\bf x})=0$
for $m\neq0$. 

Now we insert the replica-symmetric solution ansatz (\ref{AN1}) and
(\ref{AN2}) also in the other mean-field Eqs. (\ref{MMM3}) and (\ref{MF1}).
In this way we obtain in the replica limit $\mathcal{N}\to0$ the
mean-field 
\begin{eqnarray}
\Sigma({\bf x})=q({\bf x})+n_{0}({\bf x})+\lim_{\eta\rightarrow0^{+}}\sum_{m=-\infty}^{\infty}e^{i\omega_{m}\eta}\,\frac{Q_{m}({\bf x})}{\hbar\beta}\,\label{MAT1}
\end{eqnarray}
and the Gross-Pitaevskii equation
\begin{eqnarray}
\left[-\mu+2g\Sigma({\bf x})+V({\bf x})-gn_{0}({\bf x})-\frac{D}{\hbar}Q_{0}({\bf x})-\frac{\hbar^{2}}{2M}\Delta\right]\,\sqrt{n_{0}({\bf x})}=0 & ,\label{PMM}
\end{eqnarray}
 where we have set $n_{0}({\bf x})=\Psi^{*}({\bf x})\Psi({\bf x})$.

\section{Thermodynamic Properties}

Now we return to the replicated effective potential (\ref{EFF1})
and evaluate it for the special case of a $\delta$-correlated disorder
potential \eqref{Delta} and contact interaction potential \eqref{77}
at the replica-symmetric background fields \eqref{RS} and (\ref{AN1})
by taking into account Eq. (\ref{AN2}). Thus, the replicated effective
potential decomposes according to $V_{{\rm eff}}^{(\mathcal{N})}=V_{{\rm eff}}^{(\mathcal{N},1)}+V_{{\rm eff}}^{(\mathcal{N},2)}$.
The first term reads

\begin{eqnarray}
V_{{\rm eff}}^{(\mathcal{N},1)} & = & \mathcal{N}\int d{\bf x}\Biggl\{-g\Sigma^{2}({\bf x})-\frac{g}{2}\Psi^{*2}({\bf x})\Psi^{2}({\bf x})+\Psi^{*}({\bf x})\left[-\mu-\frac{\hbar^{2}}{2M}{\bf \Delta}+2g\Sigma({\bf x})+V({\bf x})-\frac{D}{2\hbar}Q_{0}^{*}({\bf x})-\frac{D}{2\hbar}Q_{0}({\bf x})\right]\nonumber \\
 &  & \times\Psi({\bf x})+\frac{D}{2\hbar}\left[Q_{0}^{*}({\bf x})+Q_{0}({\bf x})\right]\Psi^{*}({\bf x})\Psi({\bf x})+\frac{D}{2\beta\hbar^{2}}\lim_{\eta\rightarrow0^{+}}\sum_{m=-\infty}^{\infty}e^{i\omega_{m}\eta}\,\left[Q_{m}^{*}({\bf x})+Q_{m}({\bf x})\right]q_{m}({\bf x})\nonumber \\
 &  & +\frac{D}{2\hbar^{2}\beta}\,\lim_{\eta\rightarrow0^{+}}\sum_{m=-\infty}^{\infty}e^{i\omega_{m}\eta}\, Q_{m}({\bf x})Q_{-m}^{*}({\bf x})\Biggr\}+\frac{\mathcal{N}^{2}\beta D}{2}\int d{\bf x}\Biggl\{\left[\Sigma({\bf x})-\Psi^{*}({\bf x})\Psi({\bf x})\right]^{2}\nonumber \\
 &  & +\frac{1}{\left(\beta\hbar\right)^{2}}\lim_{\eta\rightarrow0^{+}}\sum_{m=-\infty}^{\infty}e^{i\omega_{m}\eta}\, q_{m}({\bf x})q_{-m}^{*}({\bf x})-\Psi^{*2}({\bf x})\Psi^{2}({\bf x})\Biggr\}\,,
\end{eqnarray}
where, again, the normal ordering is explicitly emphasized and the
second term is given by the tracelog of the integral kernel (\ref{2.44}):
\begin{eqnarray}
V_{{\rm eff}}^{(\mathcal{N},2)}=\frac{1}{2\beta}\,\mbox{Tr}\,\ln G^{-1}\,.\label{N1}
\end{eqnarray}
With the help of the Fourier-Matsubara transformation (\ref{FMT})
the latter reduces to 
\begin{eqnarray}
V_{{\rm eff}}^{(\mathcal{N},2)}=\frac{1}{2\beta}\int d{\bf x}\,\int\frac{d{\bf k}}{(2\pi)^{n}}\lim_{\eta\rightarrow0^{+}}\sum_{m=-\infty}^{\infty}e^{i\omega_{m}\eta}\ln\left[\mbox{Det}\, G_{\alpha\alpha'}^{-1}({\bf k},\omega_{m},{\bf x})\right]\,,\label{N2}
\end{eqnarray}
where the determinant of the matrix (\ref{DEKO4}) yields 
\begin{eqnarray}
\mbox{Det}\, G_{\alpha\alpha'}^{-1}({\bf k},\omega_{m},{\bf x}) & = & \left[a({\bf k},\omega_{m},{\bf x})a^{*}({\bf k},\omega_{m},{\bf x})\right]^{\mathcal{N}-1}\left[a({\bf k},\omega_{m},{\bf x})+\mathcal{N}b({\bf k},\omega_{m},{\bf x})\right]\left[a^{*}({\bf k},\omega_{m},{\bf x})+\mathcal{N}b^{*}({\bf k},\omega_{m},{\bf x})\right]\,.\label{DET}
\end{eqnarray}
Performing the replica limit $\mathcal{N}\to0$, the respective contributions
to the replicated effective potential reduce to 
\begin{eqnarray}
V_{{\rm eff}}^{(1)} & = & \lim_{\mathcal{N}\to0}\frac{V_{{\rm eff}}^{(\mathcal{N},1)}}{\mathcal{N}}=\int d{\bf x}\Biggl\{-g\Sigma^{2}({\bf x})-\frac{g}{2}\Psi^{*2}({\bf x})\Psi^{2}({\bf x})\nonumber \\
 &  & +\Psi^{*}({\bf x})\left[-\mu-\frac{\hbar^{2}}{2M}{\bf \Delta}+2g\Sigma({\bf x})+V\left({\bf x}\right)-\frac{D}{2\hbar}Q_{0}^{*}({\bf x})-\frac{D}{2\hbar}Q_{0}({\bf x})\right]\Psi({\bf x})\nonumber \\
 &  & \left.+\frac{D}{2\hbar^{2}\beta}\,\lim_{\eta\rightarrow0^{+}}\sum_{m=-\infty}^{\infty}e^{i\omega_{m}\eta}\, Q_{m}({\bf x})Q_{-m}^{*}({\bf x})+\frac{D}{2\hbar}\left[Q_{0}^{*}({\bf x})+Q_{0}({\bf x})\right]\left[q({\bf x})+\Psi^{*}({\bf x})\Psi({\bf x})\right]\right\} ,\label{VEF1}
\end{eqnarray}
and 
\begin{eqnarray}
 & V_{{\rm eff}}^{(2)}= & \lim_{\mathcal{N}\to0}\frac{V_{{\rm eff}}^{(\mathcal{N},2)}}{\mathcal{N}}=\frac{1}{2\beta}\int d{\bf x}\,\int\frac{d\mathbf{k}}{(2\pi)^{n}}\nonumber \\
 &  & \left\{ \lim_{\eta\rightarrow0^{+}}\sum_{m=-\infty}^{\infty}e^{i\omega_{m}\eta}\ln\left[-i\hbar\omega_{m}+\epsilon({\bf k})-\mu+2g\Sigma({\bf x})+V({\bf x})-\frac{D}{\hbar}Q_{m}({\bf x})\right]\right.\nonumber \\
 &  & +\lim_{\eta\rightarrow0^{+}}\sum_{m=-\infty}^{\infty}e^{i\omega_{m}\eta}\ln\left[i\hbar\omega_{m}+\epsilon({\bf k})-\mu+2g\Sigma({\bf x})+V({\bf x})-\frac{D}{\hbar}Q_{m}^{*}({\bf x})\right]\nonumber \\
 &  & \left.-\frac{\beta D\left[q({\bf x})+\Psi^{*}({\bf x})\Psi({\bf x})\right]}{\epsilon({\bf k})-\mu+2g\Sigma({\bf x})+V({\bf x})-\frac{D}{\hbar}Q_{0}({\bf x})}-\frac{\beta D\left[q({\bf x})+\Psi^{*}({\bf x})\Psi({\bf x})\right]}{\epsilon({\bf k})-\mu+2g\Sigma({\bf x})+V({\bf x})-\frac{D}{\hbar}Q_{0}^{*}({\bf x})}\right\} \,,\label{VEF2}
\end{eqnarray}
where we have inserted Eqs. (\ref{74}), \eqref{AB-2} and (\ref{DET})
into Eq. (\ref{N2}). The remaining ${\bf k}$-integrals of the logarithmic
functions in Eq. (\ref{VEF2}) are UV-divergent in all dimensions,
while the ${\bf k}$-integrals of the third and the fourth term diverge
in two and three dimensions and converge only in one dimension. Thus,
we evaluate Eq. (\ref{VEF2}) by using, again, the Schwinger integral
(\ref{SCH}) and the corresponding Schwinger representation of the
logarithm: 
\begin{eqnarray}
\ln a=-\frac{\partial}{\partial x}\,\left.\left[\frac{1}{\Gamma(x)}\int_{0}^{\infty}ds\, s^{x-1}\, e^{-as}\right]\right|_{x=0}\,.
\end{eqnarray}
With this we obtain: 
\begin{eqnarray}
V_{{\rm eff}}^{(2)} & = & -\frac{1}{2\beta}\,\left(\frac{M}{2\pi\hbar^{2}}\right)^{n/2}\lim_{\eta\rightarrow0^{+}}\sum_{m=-\infty}^{\infty}e^{i\omega_{m}\eta}\int d\mathbf{x}\Gamma\left(-\frac{n}{2}\right)\left\{ \left[-i\hbar\omega_{m}-\mu+2g\Sigma({\bf x})+V({\bf x})-\frac{D}{\hbar}Q_{m}({\bf x})\right]^{n/2}\right.\nonumber \\
 &  & \left.+\left[i\hbar\omega_{m}-\mu+2g\Sigma({\bf x})+V({\bf x})-\frac{D}{\hbar}Q_{m}^{*}({\bf x})\right]^{n/2}\right\} -\frac{D}{2}\int d\mathbf{x}\left[q({\bf x})+\Psi^{*}({\bf x})\Psi({\bf x})\right]\Gamma\left(-\frac{n}{2}+1\right)\,\left(\frac{M}{2\pi\hbar^{2}}\right)^{n/2}\nonumber \\
 &  & \times\left\{ \left[-\mu+2g\Sigma({\bf x})+V({\bf x})-\frac{D}{\hbar}Q_{0}({\bf x})\right]^{\frac{n}{2}-1}+\left[-\mu+2g\Sigma({\bf x})+V({\bf x})-\frac{D}{\hbar}Q_{0}^{*}({\bf x})\right]^{\frac{n}{2}-1}\right\} \,.\label{VEF2B}
\end{eqnarray}
As the extremum of the effective potential yields the thermodynamic
potential due to Eqs. \eqref{4} and \eqref{37}, we obtain from Eqs.
\eqref{VEF1} and (\ref{VEF2B}) the free energy:

\begin{eqnarray}
\Omega & = & \int d\mathbf{x}\left\{ -g\Sigma^{2}({\bf x})-\frac{g}{2}n_{0}^{2}({\bf x})-\sqrt{n_{0}({\bf x})}\left[\mu+\frac{\hbar^{2}}{2M}{\bf \Delta}-2g\Sigma({\bf x})-V({\bf x})+\frac{D}{\hbar}Q_{0}({\bf x})\right]\sqrt{n_{0}({\bf x})}\right.\nonumber \\
 &  & +\frac{D}{\hbar}Q_{0}({\bf x})\left[q({\bf x})+n_{0}({\bf x})\right]+\frac{D}{2\hbar^{2}\beta}\lim_{\eta\rightarrow0^{+}}\sum_{m=-\infty}^{\infty}e^{i\omega_{m}\eta}\, Q_{m}^{2}({\bf x})\nonumber \\
 &  & -D\Gamma\left(-\frac{n}{2}+1\right)\left(\frac{M}{2\pi\hbar^{2}}\right)^{n/2}\left[q({\bf x})+n_{0}({\bf x})\right]\left[-\mu+2g\Sigma({\bf x})+V({\bf x})-\frac{D}{\hbar}Q_{0}({\bf x})\right]^{\frac{n}{2}-1}\nonumber \\
 &  & \left.-\frac{1}{\beta}\,\Gamma\left(-\frac{n}{2}\right)\left(\frac{M}{2\pi\hbar^{2}}\right)^{n/2}\lim_{\eta\downarrow0}\sum_{m=-\infty}^{\infty}e^{i\omega_{m}\eta}\left[-i\hbar\omega_{m}-\mu+2g\Sigma({\bf x})+V({\bf x})-\frac{D}{\hbar}Q_{m}({\bf x})\right]^{n/2}\right\} .\label{F1}
\end{eqnarray}
Note that the particle density $n({\bf x})$, which is defined from
the expression $N=-\frac{\partial\Omega}{\partial\mu}=\int d{\bf x}n({\bf x}),$
with the particle number $N$, turns out to coincide with the mean-field
$\Sigma({\bf x})$ due to Eqs. \eqref{Qg}, \eqref{qg}, and (\ref{MAT1}):
\begin{eqnarray}
\Sigma({\bf x})=n({\bf x}).\label{QN}
\end{eqnarray}
Furthermore, all self-consistency equations \eqref{Qg}--\eqref{PMM}
can be directly obtained by extremising the thermodynamic potential
\eqref{F1} with respect to its variables $Q_{m\neq0}({\bf x})$,
$Q_{0}({\bf x})$, $q({\bf x})$, and $\sqrt{n_{0}({\bf x})}$. Indeed
the combination of the two extremisations $\frac{\delta\Omega}{\delta Q_{m\neq0}({\bf x}')}=0$
and $\frac{\delta\Omega}{\delta q({\bf x}')}=0$ gives us Eq. \eqref{Qg},
while the extremisations $\frac{\delta\Omega}{\delta Q_{0}({\bf x}')}=0$
and $\frac{\delta\Omega}{\delta\sqrt{n_{0}({\bf x}')}}=0$ yield Eqs.
\eqref{qg} and \eqref{PMM}, respectively.

Now we apply our theory, which is formulated for a general $n$-dimensional
homogeneous system, first to the three-dimensional dirty bosons, since
this case turns out to be simpler, and then to the one-dimensional
dirty bosons. The two-dimensional case cannot be treated using the
actual form of the theory as is discussed in detail below Eq. \eqref{qm-1}.

\section{Application of Hartree-Fock Mean-Field Theory in 3D}

Here we are interested in obtaining the free energy as well as the
self-consistency equations of the three-dimensional dirty boson system.
To this end, we deduce first the corresponding Matsubara coefficients.

\subsection{Matsubara Coefficients}

In three dimensions $\left(n=3\right)$, Eqs. \eqref{Q-1} and \eqref{qm-1}
reduce after performing the replica limit $\mathcal{N}\to0$ to:

\begin{equation}
Q_{m}({\bf x})=-2\sqrt{\pi}\hbar\left(\frac{M}{2\pi\hbar^{2}}\right)^{3/2}\sqrt{-i\hbar\omega_{m}-\mu+2gn({\bf x})+V({\bf x})-\frac{D}{\hbar}Q_{m}({\bf x})}\,,\label{Q}
\end{equation}
\begin{alignat}{1}
q_{m}({\bf x})= & \sqrt{\pi}D\,\left(\frac{M}{2\pi\hbar^{2}}\right)^{3/2}\frac{\left[q_{m}({\bf x})+\hbar\beta\Psi^{*}({\bf x})\Psi({\bf x})\delta_{m,0}\right]}{\sqrt{-i\hbar\omega_{m}-\mu+2g\Sigma({\bf x})+V({\bf x})-\frac{D}{\hbar}Q_{0}({\bf x})}}.\label{qm}
\end{alignat}
Equation (\ref{Q}) represents a quadratic equation for the corresponding
Matsubara coefficients $Q_{m}({\bf x})$, which is solved by:

\begin{alignat}{1}
Q_{m}({\bf x})= & -2\pi\hbar D\left(\frac{M}{2\pi\hbar^{2}}\right)^{3}\pm2\sqrt{\pi}\hbar\left(\frac{M}{2\pi\hbar^{2}}\right)^{3/2}\sqrt{-i\hbar\omega_{m}-\mu+2gn({\bf x})+V({\bf x})+\pi D^{2}\left(\frac{M}{2\pi\hbar^{2}}\right)^{3}}\,.\label{QM}
\end{alignat}
Now, we treat both cases ($m=0$ and $m\neq0$) separately.

At first, we consider the case $m=0$ and note that $Q_{0}({\bf x})$
has to be real according to Eq. (\ref{Q}). For $m=0$ Eq. \eqref{QM}
reduces to 
\begin{eqnarray}
Q_{0}({\bf x})=\left\{ \begin{array}{@{}cc}
-2\sqrt{\pi}\hbar{\displaystyle \left(\frac{M}{2\pi\hbar^{2}}\right)^{3/2}}\left[\sqrt{\pi}D{\displaystyle \left(\frac{M}{2\pi\hbar^{2}}\right)^{3/2}}+\sqrt{-\mu_{r}({\bf x})}\right]\,; & \mu_{r}({\bf x})\leq0,\\[4mm]
-2\sqrt{\pi}\hbar{\displaystyle \left(\frac{M}{2\pi\hbar^{2}}\right)^{3/2}}\left[\sqrt{\pi}D{\displaystyle \left(\frac{M}{2\pi\hbar^{2}}\right)^{3/2}}-\sqrt{-\mu_{r}({\bf x})}\right]\,; & \mu_{r}^{{\rm (crit)}}\leq\mu_{r}({\bf x})\leq0\,,
\end{array}\right.\label{Q0}
\end{eqnarray}
where we have introduced the renormalized chemical potential: 
\begin{eqnarray}
\mu_{r}({\bf x})=\mu-V({\bf x})-2gn({\bf x})-\pi D^{2}\left(\frac{M}{2\pi\hbar^{2}}\right)^{3},\label{MR}
\end{eqnarray}
and the critical chemical potential is defined by $\mu_{r}^{{\rm (crit)}}=-\pi D^{2}\left(\frac{M}{2\pi\hbar^{2}}\right)^{3}.$
Since $\mu_{r}^{{\rm (crit)}}\leq0$, we obtain from Eq. (\ref{Q0})
that the condition $\mu_{r}({\bf x})\leq0$ has to be fulfilled. 

Now, we consider the case $m\neq0$, where Eqs. (\ref{QM}) and (\ref{Q0})
are only compatible for the lower sign, i.e., we conclude: 
\begin{eqnarray}
Q_{m}({\bf x})=-2\pi\hbar D\left(\frac{M}{2\pi\hbar^{2}}\right)^{3}-2\sqrt{\pi}\hbar\left(\frac{M}{2\pi\hbar^{2}}\right)^{3/2}\sqrt{-i\hbar\omega_{m}-\mu_{r}({\bf x})}\,,\hspace*{1cm}m\neq0\,.\label{QMNN}
\end{eqnarray}
From Eq. \eqref{qm} we conclude that $q_{0}({\bf x})=\hbar\beta q({\bf x})$
has also to be real, where $q({\bf x})$ satisfies the algebraic equation:
\begin{eqnarray}
q({\bf x})=\left\{ \begin{array}{@{}cc}
\sqrt{\pi}D\left(\frac{M}{2\pi\hbar^{2}}\right)^{3/2}\frac{n_{0}({\bf x})}{\sqrt{-\mu_{r}({\bf x})}}\,; & \mu_{r}({\bf x})\leq0,\\[4mm]
-\sqrt{\pi}D\left(\frac{M}{2\pi\hbar^{2}}\right)^{3/2}\frac{n_{0}({\bf x})}{\sqrt{-\mu_{r}({\bf x})}}\,; & \mu_{r}^{{\rm (crit)}}\leq\mu_{r}({\bf x})\leq0,
\end{array}\right.\label{q}
\end{eqnarray}
and for $m\neq0$ we have $q_{m}({\bf x})=0$. At the end of Appendix
B it is shown that $q({\bf x})$ is a density and this has to be positive,
so the negative solution in Eq. \eqref{q} can be rejected. Finally,
we obtain 
\begin{equation}
q_{m}({\bf x})=\begin{cases}
0; & m\neq0,\\
\hbar\beta\sqrt{\pi}D\left(\frac{M}{2\pi\hbar^{2}}\right)^{3/2}\frac{n_{0}({\bf x})}{\sqrt{-\mu_{r}({\bf x})}}; & m=0,
\end{cases}\label{m0}
\end{equation}
Note that, due to the assumed homogeneity in time we had to put $q({\bf x}-{\bf x'},\frac{{\bf x}+{\bf x'}}{2},\tau-\tau')$
in Eq. \eqref{AN1} to be time-dependent, but according to Eq \eqref{m0}
this quantity turns out to be time-independent.

\subsection{Particle Density}

Taking into account Eqs. (\ref{Q0}) and (\ref{QMNN}), we get from
Eqs. \eqref{MAT1} and \eqref{QN} for the particle density: 
\begin{eqnarray}
n({\bf x}) & = & q({\bf x})+n_{0}({\bf x})+\frac{\Delta Q_{0}({\bf x})}{\hbar\beta}\nonumber \\
 &  & -\frac{2\sqrt{\pi}}{\beta}\left(\frac{M}{2\pi\hbar^{2}}\right)^{3/2}\lim_{\eta\rightarrow0^{+}}\sum_{m=-\infty}^{\infty}e^{i\omega_{m}\eta}\left[\sqrt{\pi}D\left(\frac{M}{2\pi\hbar^{2}}\right)^{3/2}+\sqrt{-i\hbar\omega_{m}-\mu_{r}({\bf x})}\right]\,,\label{MAT2}
\end{eqnarray}
where the following abbreviation has been introduced: 
\begin{eqnarray}
\Delta Q_{0}({\bf x})=Q_{0}({\bf x})-\lim_{m\to0}Q_{m}({\bf x})=\left\{ \begin{array}{@{}cc}
0\,; & \mu_{r}({\bf x})\leq0\\[4mm]
4\sqrt{\pi}\hbar{\displaystyle \left(\frac{M}{2\pi\hbar^{2}}\right)^{3/2}}\sqrt{-\mu_{r}({\bf x})}\,; & \mu_{r}^{{\rm (crit)}}\leq\mu_{r}({\bf x})\leq0.
\end{array}\right.\label{DQ0}
\end{eqnarray}
Then the remaining Matsubara sums (\ref{MAT2}) are evaluated by using
the zeta-function regularization method \cite{Kleinert2}. The first
sum in Eq. (\ref{MAT2}) vanishes immediately due to the Poisson formula:
\begin{eqnarray}
\sum_{m=-\infty}^{\infty}\delta(x-m)=\sum_{n=-\infty}^{\infty}e^{-2\pi inx}\,.\label{PF}
\end{eqnarray}
In order to calculate the second sum in Eq. (\ref{MAT2}), we apply
both the Schwinger integral (\ref{SCH}) and the Poisson formula (\ref{PF})
to obtain: 
\begin{eqnarray}
\lim_{\eta\rightarrow0^{+}}\sum_{m=-\infty}^{\infty}e^{i\omega_{m}\eta}\left(-i\hbar\omega_{m}+a\right)^{\nu}=\frac{\zeta_{\nu+1}\left(e^{-a\beta}\right)}{\beta^{\nu}\,\Gamma(-\nu)},\label{SUM}
\end{eqnarray}
with the polylogarithmic function $\zeta_{\nu}(z)=\sum_{\mathtt{n}=1}^{\infty}\frac{z^{\mathtt{n}}}{\mathtt{n}^{\nu}}\,.$
Thus, we obtain for the particle density 
\begin{eqnarray}
n({\bf x})=q({\bf x})+n_{0}({\bf x})+\frac{\Delta Q_{0}({\bf x})}{\hbar\beta}+\left(\frac{M}{2\pi\hbar^{2}\beta}\right)^{3/2}\,\zeta_{3/2}\left(e^{\beta\mu_{r}({\bf x})}\right)\,.\label{SI}
\end{eqnarray}

\subsection{Free Energy}

The remaining Matsubara sums in the expression for the thermodynamic
potential (\ref{F1}) are evaluated in three dimensions by using,
again, the zeta-function regularization method. Taking into account
Eqs. (\ref{Q0}), (\ref{QMNN}), and (\ref{SUM}) yields 
\begin{alignat}{1}
\frac{D}{2\hbar^{2}\beta}\lim_{\eta\rightarrow0^{+}}\sum_{m=-\infty}^{\infty}e^{i\omega_{m}\eta}\, Q_{m}^{2}({\bf x})=-2\pi D^{2}\left(\frac{M}{2\pi\hbar^{2}}\right)^{3}\left(\frac{M}{2\pi\hbar^{2}\beta}\right)^{3/2}\zeta_{3/2}\left(e^{\beta\mu_{r}({\bf x})}\right)-\frac{2\pi D^{2}\Delta Q_{0}({\bf x})}{\hbar\beta}\left(\frac{M}{2\pi\hbar^{2}}\right)^{3}
\end{alignat}
and, correspondingly, 
\begin{eqnarray}
 &  & -\frac{4\sqrt{\pi}}{3\beta}\,\left(\frac{M}{2\pi\hbar^{2}}\right)^{3/2}\lim_{\eta\rightarrow0^{+}}\sum_{m=-\infty}^{\infty}e^{i\omega_{m}\eta}\left[-i\hbar\omega_{m}-\mu+2gn({\bf x})+V({\bf x})-\frac{D}{\hbar}Q_{m}({\bf x})\right]^{3/2}=-\frac{1}{\beta}\left(\frac{M}{2\pi\hbar^{2}\beta}\right)^{3/2}\\
 &  & \times\zeta_{5/2}\left(e^{\beta\mu_{r}({\bf x})}\right)+2\pi D^{2}\left(\frac{M}{2\pi\hbar^{2}}\right)^{3}\left(\frac{M}{2\pi\hbar^{2}\beta}\right)^{3/2}\zeta_{3/2}\left(e^{\beta\mu_{r}({\bf x})}\right)+\frac{2\pi D^{2}\Delta Q_{0}({\bf x})}{\hbar\beta}\left(\frac{M}{2\pi\hbar^{2}}\right)^{3}+\frac{\Delta Q_{0}^{3}({\bf x})}{24\pi\hbar^{3}\beta}\left(\frac{M}{2\pi\hbar^{2}}\right)^{-3}\,.\nonumber 
\end{eqnarray}

According to Eq. (\ref{DQ0}) we have two solution branches of our
mean-field equations for $\mu_{r}^{({\rm crit})}\leq\mu_{r}({\bf x})\leq0$,
one with $\Delta Q_{0}({\bf x})=0$ and another one with $\Delta Q_{0}({\bf x})>0$.
As the latter solution branch yields a higher thermodynamic potential,
we do no longer consider it in the following and restrict ourselves
to the case $\Delta Q_{0}({\bf x})=0$. With this and using the mean-field
Eq. (\ref{PMM}), the thermodynamic potential (\ref{F1}) is now given
in three dimensions by: 
\begin{eqnarray}
\Omega & = & \int d{\bf x}\left\{ -g\, n^{2}({\bf x})+\frac{g}{2}n_{0}^{2}({\bf x})-\frac{1}{\beta}\left(\frac{M}{2\pi\hbar^{2}\beta}\right)^{3/2}\zeta_{5/2}\left(e^{\beta\mu_{r}({\bf x})}\right)\right.+\sqrt{n_{0}({\bf x})}\Biggl\{-gn_{0}({\bf x})\label{OZWIB}\\
 &  & \left.+\left[\sqrt{\pi}D\left(\frac{M}{2\pi\hbar^{2}}\right)^{3/2}+\sqrt{-\mu+2gn({\bf x})+V({\bf x})+\pi D^{2}\left(\frac{M}{2\pi\hbar^{2}}\right)^{3}}\right]^{2}-\frac{\hbar^{2}}{2M}{\bf \Delta}\Biggr\}\sqrt{n_{0}({\bf x})}\right\} .\nonumber 
\end{eqnarray}
Furthermore, we note that the order parameter $q({\bf x})$ turns
out not to explicitly contribute to the thermodynamic potential (\ref{OZWIB}).

\subsection{Self-Consistency Equations}

Inserting $\Delta Q_{0}({\bf x})=0$ in Eq. \eqref{SI} we obtain
for the particle density $n({\bf x)}$ the fundamental decomposition

\begin{equation}
n({\bf x})=n_{0}({\bf x})+q({\bf x})+n_{{\rm th}}\left({\bf x}\right).\label{n}
\end{equation}

It contains the order parameter of the superfluid $n_{0}({\bf x)}$,
which represents the density of the particles in the condensate, the
order parameter of the Bose-glass phase $q({\bf x)}$, which stands
for the density of the particles in the respective minima of the disorder
potential and vanishes in absence of disorder, and the thermal component
$n_{{\rm th}}\left({\bf x}\right)$ which vanishes in case of zero
temperature. Note that both order parameters $n_{0}({\bf x})$ and
$q({\bf x})$ are related to correlation functions, as is elucidated
in Appendix B. The resulting self-consistency equations for $n_{0}({\bf x})$,
$q({\bf x})$, and $n_{{\rm th}}\left({\bf x}\right)$ follow from
inserting Eq. \eqref{QN} into expressions \eqref{PMM}, \eqref{m0},
and \eqref{SI}:

\begin{equation}
\left\{ -gn{}_{0}({\bf x})+\left[\sqrt{-\mu+d^{2}+2gn({\bf x})+V({\bf x})}+d\right]^{2}-\frac{\hslash^{2}}{2M}\Delta\right\} \sqrt{n{}_{0}({\bf x})}=0,\label{6}
\end{equation}

\begin{equation}
q({\bf x})=\frac{dn{}_{0}({\bf x})}{\sqrt{-\mu+d^{2}+2gn({\bf x})+V({\bf x})}},\label{5}
\end{equation}
\begin{equation}
n_{{\rm th}}\left({\bf x}\right)=\left(\frac{M}{2\pi\beta\hslash^{2}}\right)^{3/2}\varsigma_{\ 3/2}\left(e^{\beta\ [\mu-d^{2}-2gn({\bf x})-V({\bf x})]}\right),\label{3}
\end{equation}
where $d=\sqrt{\pi}D\left(M/2\pi\hslash^{2}\right)^{3/2}$characterizes
the disorder strength. For physical reasons it is plausible to assume
that particles accumulate in the center of the trap. Thus, the differential
self-consistency Eq. \eqref{6} has to be solved with the boundary
conditions $\frac{\partial n\left(\mathbf{x}\right)}{\partial{\bf x}}\arrowvert_{{\bf x}=0}=0$
and $\frac{\partial n_{0}\left(\mathbf{x}\right)}{\partial{\bf x}}\arrowvert_{{\bf x}=0}=0$,
and the normalization condition

\begin{equation}
N=\int d\mathbf{x}n({\bf x}),\label{7}
\end{equation}
 In total we have four coupled equations, among them three algebraic
Eqs. \eqref{n}, \eqref{5}, and \eqref{3}, and one partial differential
Eq. \eqref{6}. In the absence of disorder, i.e., $d=0$, the Bose-glass
order parameter vanishes and Eq. \eqref{6} reduces to the Hartree-Fock
Gross-Pitaevskii equation in the clean case.

Note that those self-consistency Eqs. \eqref{n}--\eqref{3} can
be also obtained in a different way. To this end we rewrite the thermodynamic
potential \eqref{OZWIB} as a function of the chemical potential $\mu$,
the condensate density $n{}_{0}({\bf x})$, the Bose-glass order parameter
$q({\bf x})$ and the thermal density $n_{{\rm th}}\left({\bf x}\right)$:

\begin{eqnarray}
\Omega & = & \int d{\bf x}\Biggl\{-g\,\left[n_{0}({\bf x})+q({\bf x})+n_{{\rm th}}\left({\bf x}\right)\right]^{2}+\frac{g}{2}n_{0}^{2}({\bf x})-\frac{1}{\beta}\left(\frac{M}{2\pi\hbar^{2}\beta}\right)^{3/2}\zeta_{5/2}\left(e^{\beta\left\{ \mu-2g\left[n_{0}({\bf x})+q({\bf x})+n_{{\rm th}}\left({\bf x}\right)\right]-V({\bf x})+d^{2}\right\} }\right)\nonumber \\
 &  & +\sqrt{n_{0}({\bf x})}\Biggl\{-gn_{0}({\bf x})+\left[d+\sqrt{-\mu+2g\left[n_{0}({\bf x})+q({\bf x})+n_{{\rm th}}\left({\bf x}\right)\right]+V({\bf x})-d^{2}}\right]^{2}-\frac{\hbar^{2}}{2M}{\bf \Delta}\Biggr\}\sqrt{n_{0}({\bf x})}\Biggr\}.\label{OZWIB-1}
\end{eqnarray}

Performing a partial derivative with respect to $\mu$ and extremising
with respect to the condensate density, the Bose-glass order parameter
and the thermal density, i.e, $-\frac{\partial\Omega}{\partial\mu}=N,$
$\frac{\delta\Omega}{\delta n{}_{0}({\bf x}')}=0,$ $\frac{\delta\Omega}{\delta q({\bf x}')}=0,$
and $\frac{\delta\Omega}{\delta n_{{\rm th}}({\bf x}')}=0$, we reproduce,
indeed, Eqs. \eqref{n}--\eqref{3}. Thus, we recognize that in our
Hartree-Fock mean-field theory the order parameters can be considered
as variational parameters. This allows, in principle, to use a variational
solution method based on the principle that, among all possible configurations
of a physical system, the one that extremises some specified quantity
is realized. This method is used in physics both for theory construction
and for calculational purposes ( see, for instance, the successful
variational perturbation theory worked out in Refs.~\cite{Kleinert1,Kleinert2,Var1,Var2}).

\section{Application of Hartree-Fock Mean-Field Theory in 1D }

Now we turn to the one-dimensional case, i.e., $n=1$, where Eqs.
\eqref{Qg}--\eqref{PMM} and \eqref{QN} reduce to 
\begin{equation}
Q_{m}(x)=\sqrt{\frac{M}{2}}\frac{1}{\sqrt{-i\hbar\omega_{m}-\mu+2gn(x)+V(x)-\frac{D}{\hbar}Q_{m}(x)}},\label{Q1d}
\end{equation}

\begin{equation}
q(x)=\frac{D}{\hbar M}Q_{0}^{3}(x)\,\frac{n_{0}(x)}{1-\frac{D}{\hbar M}Q_{0}^{3}(x)},\label{q1d}
\end{equation}
the Gross-Pitaevskii equation 
\begin{eqnarray}
\left[-\mu+2gn(x)+V(x)-gn_{0}(x)-\frac{D}{\hbar}Q_{0}(x)-\frac{\hbar^{2}}{2M}\frac{\partial^{2}}{\partial x^{2}}\right]\,\sqrt{n_{0}(x)}=0,\label{PMM-1}
\end{eqnarray}
and the particle density equation 
\begin{eqnarray}
n(x)=q(x)+n_{0}(x)+\lim_{\eta\rightarrow0^{+}}\sum_{m=-\infty}^{\infty}e^{i\omega_{m}\eta}\,\frac{Q_{m}(x)}{\hbar\beta}\,.\label{MAT1-1}
\end{eqnarray}
Equation \eqref{Q1d} represents a cubic equation with respect to
$Q_{m}(x)$: 
\begin{equation}
-\frac{D}{\hbar}Q_{m}^{3}(x)+\left[-i\hbar\omega_{m}-\mu+2gn(x)+V(x)\right]Q_{m}^{2}(x)-\frac{M}{2}=0,\label{Q3}
\end{equation}
whose solution should be inserted into Eqs. \eqref{q1d}--\eqref{Q3}.
To this end we have to use the Cardan method \cite{Greiner}, which
is characterized by a discriminant. For the sake of simplicity, we
restrict ourselves to the zero temperature case, where only the $m=0$
term contributes. In this case the discriminant has a real value and
the Cardan method can be applied. According to the sign of the discriminant
$\delta_{0}$ we get the following real solutions for $Q_{0}(x)$:

\begin{eqnarray}
Q_{0}(x)=\left\{ \begin{array}{@{}cc}
\!\!\!\!\!\!\!\!\!\!\!\!\!\!\!\!\!\!\!\!\!\!\!\!\!\!\!\sqrt[3]{\frac{-p+\sqrt{\delta_{0}}}{2}}+\sqrt[3]{\frac{-p-\sqrt{\delta_{0}}}{2}}+\frac{\hbar}{3D}\left[-\mu+2gn(x)+V(x)\right]\,; & \delta_{0}>0,\\[4mm]
\!\!\!\!\!\!\!\!\!\!\!\!\!\!\!\!\!\!\!\!\!\!\!\!\!\!\sqrt[3]{\frac{-p+i\sqrt{-\delta_{0}}}{2}}+\sqrt[3]{\frac{-p-i\sqrt{-\delta_{0}}}{2}}+\frac{\hbar}{3D}\left[-\mu+2gn(x)+V(x)\right]; & \delta_{0}\leq0,\\
e^{\pm\frac{2i\pi}{3}}\sqrt[3]{\frac{-p+i\sqrt{-\delta_{0}}}{2}}+e^{\mp\frac{2i\pi}{3}}\sqrt[3]{\frac{-p-i\sqrt{-\delta_{0}}}{2}}+\frac{\hbar}{3D}\left[-\mu+2gn(x)+V(x)\right]; & \delta_{0}\leq0,
\end{array}\right.\label{Q0-1}
\end{eqnarray}
with the abbreviation $p=-\frac{2\hbar^{3}}{27D^{3}}\left[-\mu+2gn(x)+V(x)\right]^{3}+\frac{\hbar M}{2D}.$
The correct solution of $Q_{0}(x)$ has, according to Eq. \eqref{q1d},
to be positive and can only be selected after choosing the form of
the trap and by ensuring a minimal free energy.

At zero temperature Eqs. \eqref{q1d} and \eqref{PMM-1} remain the
same, but Eq. \eqref{MAT1-1} reduces to: 
\begin{eqnarray}
n(x) & = & q(x)+n_{0}(x),\label{145}
\end{eqnarray}
and the free energy \eqref{F1} specializes, with \eqref{Q1d}, to:
\begin{eqnarray}
\Omega & = & \int dx\Biggl\{-g\left[n_{0}(x)+q(x)\right]^{2}-\frac{g}{2}n_{0}^{2}(x)\nonumber \\
 &  & -\sqrt{n_{0}(x)}\left\{ \mu+\frac{\hbar^{2}}{2M}\frac{\partial^{2}}{\partial x^{2}}-2g\left[n_{0}(x)+q(x)\right]-V(x)+\frac{D}{\hbar}Q_{0}(x)\right\} \sqrt{n_{0}(x)}\Biggr\}.\label{F1-1}
\end{eqnarray}

After inserting Eq. \eqref{145} into Eq. \eqref{Q0-1} and then inserting
the result into the free energy expression \eqref{F1-1}, the three
self-consistency equations \eqref{q1d}, \eqref{PMM-1}, and \eqref{145}
can be directly obtained by extremising the free energy with respect
to its variables $q(x)$, $n_{0}(x)$ and $\mu$, i.e., $-\frac{\partial\Omega}{\partial\mu}=N,\,\frac{\delta\Omega}{\delta n{}_{0}(x')}=0$
and $\frac{\delta\Omega}{\delta q(x')}=0$, respectively. So also
in one dimension our Hartree-Fock mean-field theory can be based on
identifying the order parameters as variational parameters.

\section{Conclusions and Outlook}

In this paper, we developed in detail a Hartree-Fock mean-field theory
on the basis of the replica method for a trapped delta-correlated
weakly interacting Bose gas in $n$ dimensions at finite temperature.
This allowed us to get the free energy as well as the underlying self-consistency
equations for the respective components of the particle density. In
the end, we applied this theory to one-dimensional and three-dimensional
dirty bosons. 

On the basis of these self-consistency relations the possible emergence
of a Bose-glass region in trapped quasi-1D Bose-Einstein condensed
systems in the presence of delta-correlated disorder is analyzed in
Ref.~\cite{Paper1}. Analytical calculations based on the present
Hartree-Fock mean-field theory as well as detailed numerical simulations
show unambiguously the existence of  a Bose-glass region,
whose spatial distribution turns out to change with the disorder strength.
For small disorder strengths the Bose-glass region
emerges at the edge of the atomic cloud, while in the intermediate
disorder regime it is located in the trap center. But no quantum
phase transition from the superfluid to the Bose-glass phase could
be detected neither in the weak nor in the intermediate disorder regime. 

The case of tree-dimensional trapped dirty bosons is investigated
within the Hartree-Fock mean-field theory in Ref.~\cite{Paper3},
where the existence of a first-order quantum phase transition
from the superfluid to the Bose-glass at zero temperature for a harmonically
trapped delta-correlated dirty boson is detected at a critical
disorder strength, which qualitatively agrees with findings in the
literature. At finite temperature the impact of both temperature and
disorder fluctuations on the respective components of the density
as well as their Thomas-Fermi radii are studied. In particular, we
found that a superfluid region, a Bose-glass region, and a thermal
region coexist. Furthermore, depending on the respective system parameters,
three phase transitions are detected, namely, one from the superfluid
to the Bose-glass phase, another one from the Bose-glass to the thermal
phase, and, finally, one directly from the superfluid to the thermal
phase.

We expect that the seminal results obtained in Refs.~\cite{Paper1,Paper3},
which follow from the theory worked out in this paper, are useful
for a quantitative analysis of ongoing experiments for dirty bosons
in quasi one- and three-dimensional harmonic traps. Furthermore, we
expect that the UV-divergency encountered in our two-dimensional theory
according to section VI can be eliminated within a proper renormalization
program. The resulting self-consistency equations in two dimensions
would then be suitable, for instance, to analyze the localization
properties of dirty photons in a microcavity \cite{Photons}.
This seems to be insofar a quite challenging research problem as the
superfluid to Bose-glass transition could (not) be found in 3D (1D)
on the basis of  the theory of this paper \cite{Paper1,Paper3}. Thus,
in view of the existence of the Bose-glass phase, the case of trapped
dirty photons is marginal.

It should be noted that the replica symmetry can break \cite{Intro-40}.
For instance, the so-called replica-symmetric solution of the Sherrington-Kirkpatrick
was shown to break down below a critical temperature \cite{Intro24,Intro28}.
Therefore, Parisi introduced the scheme of replica-symmetry breaking
(RSB) \cite{Intro-32,Intro-34,Intro-35,Intro-39}. It turns out to
yield a stable solution for the Sherrington-Kirkpatrick model for
all temperatures. The physical origin of RSB is the existence of many
local minima of the complicated free energy, which are separated by
high barriers. Practically one has to compare the free energies associated
with the RS and RSB solutions and verify whether the free energy of
the RSB solution is smaller. If this is the case this proofs that
RS is broken. In the case of dirty bosons it still has to be shown
whether RSB lowers the free energy or not.
\begin{acknowledgments}
The authors gratefully thank Antun Bala\v{z} and Robert Graham. Furthermore,
we acknowledge financial support from the German Academic and Exchange
Service (DAAD) and the German Research Foundation (DFG) via the Collaborative
Research Center SFB/TR49 ``Condensed Matter Systems with Variable
Many-Body Interactions''.
\end{acknowledgments}

\appendix

\section{Disorder Potential}

\begin{widetext}

\begin{figure}[t]
\noindent \centering{}\includegraphics[width=0.45\textwidth]{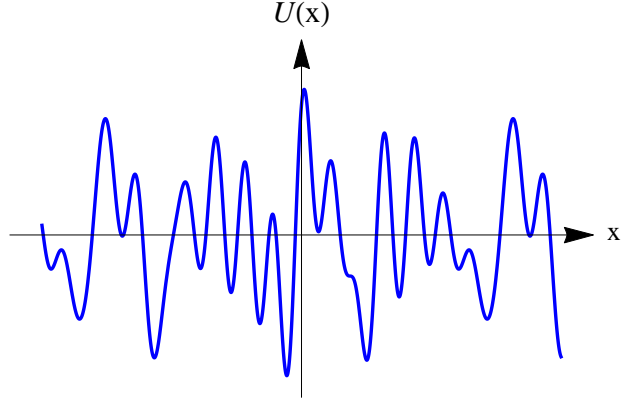}
\protect\protect\caption{\label{fig:U(X)}Example for a realization of a frozen disorder potential
$U({\bf x})$ with vanishing expectation value (\ref{ZP1}).}
\end{figure}

\end{widetext}

Here we introduce the statistical properties of the considered disorder
potential $U({\bf x})$ which fluctuates at each space point ${\bf x}$
from realization to realization (see Fig.~\ref{fig:U(X)}). Such
a frozen disorder potential serves, for instance, for modeling superfluid
helium in porous media \cite{Intro-91,Intro-92,Intro-93,Intro-94},
where the pores can be modeled by statistically distributed local
scatterers. In the following we assume for the disorder potential
that it is homogeneous after the disorder ensemble average, i.e.,
after having performed the average $\overline{\bullet}$ over all
possible realizations. Thus, the expectation value of the disorder
potential vanishes without loss of generality

\begin{eqnarray}
\overline{U({\bf x})}=0.\label{ZP1}
\end{eqnarray}
Indeed, due to the homogeneity, the disorder ensemble average $\overline{U({\bf x})}$
represents a constant, which can be absorbed without loss of generality
into the chemical potential within a grand-canonical description.
Furthermore, a homogeneous disorder potential has a correlation function
which depends on the difference of the space points: 
\begin{eqnarray}
\overline{U({\bf x}_{1})U({\bf x}_{2})}=D({\bf x}_{1}-{\bf x}_{2})\,.\label{ZP2}
\end{eqnarray}
In case of a Gaussian correlated disorder in $n$ spatial dimensions
we have 
\begin{eqnarray}
D({\bf x}_{1}-{\bf x}_{2})=D\,\frac{e^{-({\bf x}_{1}-{\bf x}_{2})^{2}/2\xi^{2}}}{(2\pi\xi^{2})^{n/2}}\,,\label{ZP3}
\end{eqnarray}
where its coherence length $\xi$ can be identified with the average
extension of the pores \cite{HM-4}. If one is not interested in a
quantitative model for interpreting experimental measurements, one
can neglect this spatial extension of the pores. In the limit of a
vanishing coherence length $\xi$ we obtain a qualitative model for
disordered bosons with a delta correlation: 
\begin{eqnarray}
D({\bf x}_{1}-{\bf x}_{2})=D\,\delta({\bf x}_{1}-{\bf x}_{2})\,.\label{ZP4}
\end{eqnarray}
Here the parameter $D$ is proportional to the density of pores and
represents a measure for the disorder strength.

As a next step we consider the probability distribution $P[U]$, which
is a functional of the disorder potential $U({\bf x})$. To this end
we define expectation values such as (\ref{ZP1}) and (\ref{ZP2})
by the functional integral: 
\begin{eqnarray}
\overline{\,\,\phantom{V}\bullet\phantom{V}\,\,}=\int{\cal D}U\,\,\bullet\,\, P[U]\,.\label{ZP5}
\end{eqnarray}
Here the functional integral stands for an infinite product of ordinary
integrals with respect to all possible values of the disorder potential
$U({\bf x})$ at all space points ${\bf x}$ \cite{Kleinert3}: 
\begin{eqnarray}
\int{\cal D}U=\prod_{{\bf x}}\,\int_{-\infty}^{\infty}dU({\bf x})\,.\label{ZP5B}
\end{eqnarray}
The functional measure has to be chosen according to 
\begin{eqnarray}
\int{\cal D}U\, P[U]=1\,,\label{ZP6}
\end{eqnarray}
so that the probability distribution is normalized: $\langle\,1\,\rangle=1$.

Provided that $P[U]$ is Gaussian distributed, it is uniquely fixed
by both expectation values (\ref{ZP1}) and (\ref{ZP2}) according
to 
\begin{eqnarray}
P[U]=\exp\left\{ -\frac{1}{2}\int d{\bf x}\int d{\bf x}'\, D^{-1}({\bf x}-{\bf x'})U({\bf x})U({\bf x'})\right\} \,,\label{ZP7}
\end{eqnarray}
where the integral kernel $D^{-1}({\bf x}-{\bf x'})$ represents the
functional inverse of the correlation function (\ref{ZP3}): 
\begin{eqnarray}
\int d{\bf x}\, D^{-1}({\bf x}_{1}-{\bf x})D({\bf x}-{\bf x}_{2})=\delta({\bf x}_{1}-{\bf x}_{2})\,.\label{ZP8}
\end{eqnarray}
For instance, we obtain for the $\delta$-correlation (\ref{ZP4})
from (\ref{ZP8}) the integral kernel: 
\begin{eqnarray}
D^{-1}({\bf x}_{1}-{\bf x}_{2})=\frac{1}{D}\,\delta({\bf x}_{1}-{\bf x}_{2})\,.\label{ZP9}
\end{eqnarray}
We are interested in calculating higher moments of the probability
distribution (\ref{ZP7}). To this end we consider the following generating
functional 
\begin{eqnarray}
I[j]=\overline{\exp\left\{ \int d{\bf x}\, j({\bf x})U({\bf x})\right\} },\label{ZP10}
\end{eqnarray}
with the auxiliary current field $j({\bf x})$ which represents according
to (\ref{ZP5}) and (\ref{ZP7}) a Gaussian functional integral with
the result \cite{Kleinert3} 
\begin{eqnarray}
I[j]=\exp\left\{ \frac{1}{2}\int d{\bf x}\int d{\bf x}'\, D({\bf x}-{\bf x'})j({\bf x})j({\bf x'})\right\}  & .\label{**}
\end{eqnarray}
The respective moments of the probability distribution (\ref{ZP7})
follow from successive functional derivatives of the generating functional
(\ref{ZP10}) with respect to the auxiliary current field $j({\bf x})$.
Indeed, we obtain for the first two moments: 
\begin{eqnarray}
\overline{U({\bf x}_{1})} & = & \left.\frac{\delta I[j]}{\delta j({\bf x}_{1})}\right|_{j({\bf x})=0}\,,\label{ZP15}\\
\overline{U({\bf x}_{1})U({\bf x}_{2})} & = & \left.\frac{\delta^{2}I[j]}{\delta j({\bf x}_{1})\delta j({\bf x}_{2})}\right|_{j({\bf x})=0}\,.\label{ZP16}
\end{eqnarray}
Inserting (\ref{**}) into (\ref{ZP15}) and (\ref{ZP16}) leads then,
indeed, to (\ref{ZP1}) and (\ref{ZP2}). In a similar way also higher
correlation functions are evaluated. Whereas the expectation values
of all odd products of disorder potentials vanish, those with an even
product are evaluated according to the Wick rule. So we obtain, for
instance: 
\begin{eqnarray}
\overline{U({\bf x}_{1})U({\bf x}_{2})U({\bf x}_{3})U({\bf x}_{4})} & = & D({\bf x}_{1}-{\bf x}_{2})D({\bf x}_{3}-{\bf x}_{4})+D({\bf x}_{1}-{\bf x}_{3})D({\bf x}_{2}-{\bf x}_{4})+D({\bf x}_{1}-{\bf x}_{4})D({\bf x}_{2}-{\bf x}_{3})\,.\label{ZP17}
\end{eqnarray}

In the case that the probability distribution $P[U]$ is not Gaussian,
its generating functional \eqref{ZP10} contains more than the second
cumulant \cite{Risken}, so we have as a straight-forward generalization
of \eqref{**}: 
\begin{eqnarray}
I[j]=\exp\left\{ \sum_{i=2}^{\infty}\frac{(-1)^{i-1}}{i!}\int d{\bf x}_{1}\cdots\int d{\bf x}_{i}\, D^{\left(i\right)}(\mathbf{x}_{1},\ldots,\mathbf{x}_{i})j({\bf x}_{1})\cdots j({\bf x}_{i})\right\} \,,\label{ZP12}
\end{eqnarray}
where $D^{\left(i\right)}(\mathbf{x}_{1},\ldots,\mathbf{x}_{i})$
denotes the i$^{{\rm {th}}}$ cumulant. Indeed, Eq. \eqref{ZP12}
reduces with $D^{\left(2\right)}(\mathbf{x}_{1},\mathbf{x}_{2})=D(\mathbf{x}_{1},\mathbf{x}_{2})$
and $D^{\left(i\right)}(\mathbf{x}_{1},\ldots,\mathbf{x}_{i})=0$
for $i\geq3$ to Eq. \eqref{**}.

\section{Correlation Functions and Order Parameters}

In the following we fix the physical interpretation of the two order
parameters $n_{0}({\bf x})$ and $q({\bf x})$ that our mean-field
theory contains. To this end we follow the notion of classical and
quantum spin-glass theory \cite{Intro-101,Intro-35,Hertz} and investigate
how these quantities are related to correlation functions.

We start with considering the grand-canonical average of the Bose
field: 
\begin{eqnarray}
\langle\psi({\bf x},\tau)\rangle=\frac{1}{{\cal Z}}\,\oint{\cal D}\psi^{*}\oint{\cal D}\psi\,\psi({\bf x},\tau)\, e^{-{\cal A}[\psi^{*},\psi]/\hbar}\,,
\end{eqnarray}
which represents a functional of the disorder potential $U({\bf x})$
due to the action (\ref{eq:A}). In order to evaluate its disorder
expectation value we apply again the replica method. To this end we
identify $\psi({\bf x},\tau)$ with $\psi_{\alpha}({\bf x},\tau)$
and add further $\mathcal{N}-1$ Bose fields according to: 
\begin{eqnarray}
\langle\psi({\bf x},\tau)\rangle=\frac{1}{{\cal Z}^{\mathcal{N}}}\left\{ \prod_{\alpha'=1}^{N}\oint{\cal D}\psi_{\alpha'}^{*}\oint{\cal D}\psi_{\alpha'}\right\} \,\psi_{\alpha}({\bf x},\tau)\,\exp\left\{ -\frac{1}{\hbar}\sum_{\alpha'=1}^{\mathcal{N}}{\cal A}[\psi_{\alpha'}^{*},\psi_{\alpha'}]\right\} \,.
\end{eqnarray}
As the right-hand side is independent of the replica index $\alpha$,
we obtain in the replica limit $\mathcal{N}\to0$: 
\begin{eqnarray}
\langle\psi({\bf x},\tau)\rangle=\lim_{\mathcal{N}\to0}\frac{1}{\mathcal{N}}\sum_{\alpha=1}^{\mathcal{N}}\left\{ \prod_{\alpha'=1}^{\mathcal{N}}\oint{\cal D}\psi_{\alpha'}^{*}\oint{\cal D}\psi_{\alpha'}\right\} \,\psi_{\alpha}({\bf x},\tau)\,\exp\left\{ -\frac{1}{\hbar}\sum_{\alpha'=1}^{\mathcal{N}}{\cal A}[\psi_{\alpha'}^{*},\psi_{\alpha'}]\right\} \,.
\end{eqnarray}
Now we are in a position to perform the averaging with respect to
the disorder potential $U({\bf x})$ by applying again the generating
functional (\ref{ZP12}) with the auxiliary current field (\ref{ZP27}).
Thus we obtain the following replica representation of the grand-canonical
average of the Bose field: 
\begin{eqnarray}
\overline{\langle\psi({\bf x},\tau)\rangle}=\lim_{\mathcal{N}\to0}\frac{1}{\mathcal{N}}\sum_{\alpha=1}^{\mathcal{N}}\left\{ \prod_{\alpha'=1}^{\mathcal{N}}\oint{\cal D}\psi_{\alpha'}^{*}\oint{\cal D}\psi_{\alpha'}\right\} \,\psi_{\alpha}({\bf x},\tau)\, e^{-{\cal A}^{(\mathcal{N})}[\psi^{*},\psi]/\hbar}\label{1P}
\end{eqnarray}
with the replica action (\ref{28}) as we restrict ourselves also
here to the second cumulant. In a similar way we yield for the two-point
function: 
\begin{eqnarray}
\overline{\langle\psi({\bf x},\tau)\psi^{*}({\bf x'},\tau')\rangle}=\lim_{\mathcal{N}\to0}\frac{1}{\mathcal{N}}\sum_{\alpha=1}^{\mathcal{N}}\left\{ \prod_{\alpha'=1}^{\mathcal{N}}\oint{\cal D}\psi_{\alpha'}^{*}\oint{\cal D}\psi_{\alpha'}\right\} \,\psi_{\alpha}({\bf x},\tau)\psi_{\alpha}^{*}({\bf x'},\tau')\, e^{-{\cal A}^{(\mathcal{N})}[\psi^{*},\psi]/\hbar}\,.\label{2P}
\end{eqnarray}
In order to further evaluate $\mathrm{n}$-point functions of the
form (\ref{1P}) and (\ref{2P}), we introduce the generating functional:
\begin{eqnarray}
\overline{{\cal Z}[j^{*},j]}=\left\{ \prod_{\alpha=1}^{\mathcal{N}}\oint{\cal D}\psi_{\alpha}^{*}\oint{\cal D}\psi_{\alpha}\right\} \, e^{-{\cal A}^{(\mathcal{N})}[\psi^{*},\psi;j^{*},j]/\hbar}\,,\label{GF1}
\end{eqnarray}
where each Bose field $\psi_{\alpha}^{*}({\bf x},\tau),\psi_{\alpha}({\bf x},\tau)$
is coupled to its own current field $j_{\alpha}({\bf x},\tau),j_{\alpha}^{*}({\bf x},\tau)$
via the action:

\begin{eqnarray}
{\cal A}^{(\mathcal{N})}[\psi^{*},\psi;j^{*},j]={\cal A}^{(\mathcal{N})}[\psi^{*},\psi]-\int_{0}^{\hbar\beta}d\tau\int d{\bf x}\sum_{\alpha=1}^{\mathcal{N}}\Big\{ j_{\alpha}^{*}({\bf x},\tau)\psi_{\alpha}({\bf x},\tau)+\psi_{\alpha}^{*}({\bf x},\tau)j_{\alpha}({\bf x},\tau)\Big\}\,.\label{GF2}
\end{eqnarray}
Indeed, performing successive functional derivatives with respect
to the current fields $j_{\alpha}({\bf x},\tau),j_{\alpha}^{*}({\bf x},\tau)$,
we obtain the 1- and 2-point function (\ref{1P}) and (\ref{2P})
from the generating functional (\ref{GF1}) and (\ref{GF2}) according
to: 
\begin{eqnarray}
\overline{\langle\psi({\bf x},\tau)\rangle} & = & \left.\lim_{\mathcal{N}\to0}\frac{\hbar}{\mathcal{N}}\sum_{\alpha=1}^{\mathcal{N}}\frac{\delta\overline{{\cal Z}[j^{*},j]}}{\delta j_{\alpha}^{*}({\bf x},\tau)}\right|_{j({\bf x},\tau)=0}^{j^{*}({\bf x},\tau)=0}\,,\label{1PB}\\
\overline{\langle\psi({\bf x},\tau)\psi^{*}({\bf x'},\tau')\rangle} & = & \left.\lim_{\mathcal{N}\to0}\frac{\hbar^{2}}{\mathcal{N}}\sum_{\alpha=1}^{\mathcal{N}}\frac{\delta^{2}\overline{{\cal Z}[j^{*},j]}}{\delta j_{\alpha}^{*}({\bf x},\tau)\delta j_{\alpha}({\bf x'},\tau')}\right|_{j({\bf x},\tau)=0}^{j^{*}({\bf x},\tau)=0}\,.\label{2PB}
\end{eqnarray}
Thus, it remains to calculate the generating functional $\overline{{\cal Z}[j^{*},j]}$
within our Hartree-Fock mean-field theory. To this end we perform
the background expansions (\ref{26}) and assume again that the background
fields have the replica symmetry form (\ref{RS}), so we have: 
\begin{eqnarray}
\overline{{\cal Z}[j^{*},j]} & = & \exp\left\{ -\beta V_{{\rm eff}}^{(N)}+\frac{1}{\hbar}\int_{0}^{\hbar\beta}d\tau\int d{\bf x}\sum_{\alpha=1}^{N}\Big[j_{\alpha}^{*}({\bf x},\tau)\,\Psi({\bf x})+\Psi^{*}({\bf x})\, j_{\alpha}({\bf x},\tau)\Big]\right.\nonumber \\
 &  & \left.+\frac{1}{\hbar^{2}}\int_{0}^{\hbar\beta}d\tau\int_{0}^{\hbar\beta}d\tau'\int d{\bf x}\int d{\bf x}'\sum_{\alpha=1}^{N}\sum_{\alpha'=1}^{N}j_{\alpha}^{*}({\bf x},\tau)\,\,\langle\delta\psi_{\alpha}({\bf x},\tau)\delta\psi_{\alpha'}^{*}({\bf x'},\tau')\rangle j_{\alpha'}({\bf x'},\tau')\right\} \,.\label{EFF}
\end{eqnarray}
Inserting (\ref{EFF}) into (\ref{1PB}) and (\ref{2PB}), yields
\begin{eqnarray}
\overline{\langle\psi({\bf x},\tau)\rangle}=\sqrt{n_{0}({\bf x})}
\end{eqnarray}
and, by taking into account (\ref{EXP}): 
\begin{eqnarray}
\overline{\langle\psi({\bf x},\tau)\psi^{*}({\bf x'},\tau')\rangle} & = & \sqrt{n_{0}({\bf x})n_{0}({\bf x'})}+g_{1}\left({\bf x}-{\bf x'},\frac{{\bf x}+{\bf x'}}{2};\tau-\tau'\right)+g_{2}\left({\bf x}-{\bf x'},\frac{{\bf x}+{\bf x'}}{2};\tau-\tau'\right)\,.\label{COI}
\end{eqnarray}
Now we need just to evaluate the functions $g_{1}({\bf x}-{\bf x'},\frac{{\bf x}+{\bf x'}}{2};\tau-\tau')$
and $g_{2}({\bf x}-{\bf x'},\frac{{\bf x}+{\bf x'}}{2};\tau-\tau'),$
respectively. Inserting \eqref{74}, \eqref{AB-2} into \eqref{G1},
\eqref{G2} and using the Schwinger integral \eqref{SCH}, \cite[(3.471.9)]{Gradshteyn},
and \cite[(8.469.3)]{Gradshteyn}, as well as performing the replica
limit $\mathcal{N}\to0$ yields: 
\begin{alignat}{1}
g_{1}\left({\bf x}-{\bf x'},\frac{{\bf x}+{\bf x'}}{2};\tau-\tau'\right) & =\frac{\sqrt{\pi}}{\beta}\left(\frac{M}{2\pi\hbar^{2}}\right)^{n/2}\left(\frac{2\hbar^{2}}{M({\bf x}-{\bf x'})^{2}}\right)^{\frac{n-1}{4}}\label{ZW1}\\
 & \negthickspace\negthickspace\negthickspace\negthickspace\negthickspace\negthickspace\negthickspace\negthickspace\negthickspace\negthickspace\negthickspace\negthickspace\negthickspace\negthickspace\negthickspace\negthickspace\negthickspace\negthickspace\negthickspace\negthickspace\negthickspace\negthickspace\negthickspace\negthickspace\negthickspace\negthickspace\negthickspace\negthickspace\negthickspace\negthickspace\negthickspace\negthickspace\negthickspace\negthickspace\negthickspace\negthickspace\times\sum_{m=-\infty}^{\infty}\frac{1}{\left[-i\hbar\omega_{m}+V(\frac{{\bf x}+{\bf x'}}{2})-\mu+2g\Sigma(\frac{{\bf x}+{\bf x'}}{2})-\frac{D}{\hbar}Q_{m}(\frac{{\bf x}+{\bf x'}}{2})\right]^{\frac{3-n}{4}}}\nonumber \\
 & \negthickspace\negthickspace\negthickspace\negthickspace\negthickspace\negthickspace\negthickspace\negthickspace\negthickspace\negthickspace\negthickspace\negthickspace\negthickspace\negthickspace\negthickspace\negthickspace\negthickspace\negthickspace\negthickspace\negthickspace\negthickspace\negthickspace\negthickspace\negthickspace\negthickspace\negthickspace\negthickspace\negthickspace\negthickspace\negthickspace\negthickspace\negthickspace\negthickspace\negthickspace\negthickspace\negthickspace\times\exp\left\{ -i\omega_{m}(\tau-\tau')-\sqrt{\frac{2M}{\hbar^{2}}\left[-i\hbar\omega_{m}+V\left(\frac{{\bf x}+{\bf x'}}{2}\right)-\mu+2g\Sigma\left(\frac{{\bf x}+{\bf x'}}{2}\right)-\frac{D}{\hbar}Q_{m}\left(\frac{{\bf x}+{\bf x'}}{2}\right)\right]}\,\,|{\bf x}-{\bf x'}|\right\} \nonumber 
\end{alignat}
and 
\begin{eqnarray}
g_{2}\left({\bf x}-{\bf x'},\frac{{\bf x}+{\bf x'}}{2};\tau-\tau'\right) & = & \sqrt{\pi}D\left(\frac{M}{2\pi\hbar^{2}}\right)^{n/2}\left(\frac{2\hbar^{2}}{M({\bf x}-{\bf x'})^{2}}\right)^{\frac{n-1}{4}}\left[q\left(\frac{{\bf x}+{\bf x'}}{2}\right)+\Psi^{*}\left(\frac{{\bf x}+{\bf x'}}{2}\right)\Psi\left(\frac{{\bf x}+{\bf x'}}{2}\right)\right]\nonumber \\
 &  & \times\frac{\sqrt{\frac{M}{2\hbar^{2}}\left[V(\frac{{\bf x}+{\bf x'}}{2})-\mu+2g\Sigma(\frac{{\bf x}+{\bf x'}}{2})-\frac{D}{\hbar}Q_{0}(\frac{{\bf x}+{\bf x'}}{2})\right]}|{\bf x}-{\bf x'}|+\frac{3-n}{4}}{\left[V(\frac{{\bf x}+{\bf x'}}{2})-\mu+2g\Sigma(\frac{{\bf x}+{\bf x'}}{2})-\frac{D}{\hbar}Q_{0}(\frac{{\bf x}+{\bf x'}}{2})\right]^{\frac{7-n}{4}}}\label{ZW2}\\
 &  & \times\exp\left\{ -\sqrt{\frac{2M}{\hbar^{2}}\left[V\left(\frac{{\bf x}+{\bf x'}}{2}\right)-\mu+2g\Sigma\left(\frac{{\bf x}+{\bf x'}}{2}\right)-\frac{D}{\hbar}Q_{0}\left(\frac{{\bf x}+{\bf x'}}{2}\right)\right]}\,\,|{\bf x}-{\bf x'}|\right\} ,\nonumber 
\end{eqnarray}
respectively. Note that the function $g_{2}\left({\bf x}-{\bf x'},\frac{{\bf x}+{\bf x'}}{2};\tau-\tau'\right)$
turns out not to depend on $\tau-\tau'$ at all.

Correspondingly, we determine the disorder average of the 4-point
function $\left|\langle\psi({\bf x},\tau)\psi^{*}({\bf x'},\tau')\rangle\right|^{2}=\langle\psi({\bf x},\tau)\psi^{*}({\bf x'},\tau')\rangle\,\langle\psi^{*}({\bf x},\tau)\psi({\bf x'},\tau')\rangle,$
which has the replica representation: 
\begin{eqnarray}
\overline{\left|\langle\psi({\bf x},\tau)\psi^{*}({\bf x'},\tau')\rangle\right|^{2}}=\lim_{N\to0}\frac{\hbar^{4}}{\mathcal{N}(\mathcal{N}-1)}\,\left.\sum_{\alpha\neq\alpha'}\,\,\frac{\delta^{4}\overline{{\cal Z}[j^{*},j]}}{\delta j_{\alpha}^{*}({\bf x},\tau)j_{\alpha}({\bf x'},\tau')\delta j_{\alpha'}^{*}({\bf x'},\tau')j_{\alpha'}({\bf x},\tau)}\right|_{j({\bf x},\tau)=0}^{j^{*}({\bf x},\tau)=0}\,.\label{4P}
\end{eqnarray}
Inserting the generating functional (\ref{EFF}) into (\ref{4P})
leads to: 
\begin{eqnarray}
\overline{\left|\langle\psi({\bf x},\tau)\psi^{*}({\bf x'},\tau')\rangle\right|^{2}}=|\overline{\langle\psi({\bf x},\tau)\psi^{*}({\bf x'},\tau')\rangle}|^{2}+n_{0}({\bf x})\, g_{2}(\mathbf{0},{\bf x'};0)+n_{0}({\bf x'})\, g_{2}(\mathbf{0},{\bf x};0)+g_{2}(\mathbf{0},{\bf x};0)g_{2}(\mathbf{0},{\bf x}';0)\,.\label{4PB}
\end{eqnarray}
Now we are in the position to investigate the 2- and the 4-point function
(\ref{COI}) and (\ref{4PB}) for special values of their spatio-temporal
arguments. At first, we set $\tau=\tau'$ and study their behavior
in the long-range limit $|{\bf x}-{\bf x'}|\to\infty$. From (\ref{COI})
with (\ref{ZW1}) and (\ref{ZW2}) we obtain for the 2-point function:
\begin{eqnarray}
\lim_{|{\bf x}-{\bf x'}|\to\infty}\overline{\langle\psi({\bf x},\tau)\psi^{*}({\bf x'},\tau)\rangle}=\sqrt{n_{0}({\bf x})n_{0}({\bf x}')}\,.
\end{eqnarray}
We read off from (\ref{qg}) and (\ref{ZW2}) that $q({\bf x})=g_{2}\left(0,{\bf x};0\right)$,
so that the 4-point function (\ref{4PB}) leads to: 
\begin{eqnarray}
\lim_{|{\bf x}-{\bf x'}|\to\infty}\overline{\left|\langle\psi({\bf x},\tau)\psi^{*}({\bf x'},\tau)\rangle\right|^{2}}=\left[n_{0}({\bf x})+q({\bf x})\right]\left[n_{0}({\bf x}')+q({\bf x'})\right]\,.
\end{eqnarray}

Following the notion of classical spin-glass theory \cite{Intro-35,Hertz},
this result justifies to consider the quantities $n_{0}({\bf x})$
and $q({\bf x})$ as the order parameters of the condensate and the
Bose-glass phase, respectively. However, in analogy to quantum spin-glass
theory \cite{Intro-101}, the Bose-glass order parameter $q({\bf x})$,
which has been introduced in Ref.~\cite{Intro-90} in close analogy
to the Edward-Anderson order parameter of spin-glasses \cite{Intro-101},
should also be related to the long-time limit $|\tau-\tau'|\to\infty$
of the 2-point function (\ref{COI}) at $T=0$. At $T=0$ the term
(\ref{ZW1}) vanishes, whereas (\ref{ZW2}) remains valid as it is
temperature independent. By setting ${\bf x}={\bf x'}$, we consider
the behavior of the 2-point function (\ref{COI}) in the long-time
limit $|\tau-\tau'|\to\infty$ and read off from (\ref{qg}), (\ref{COI})--(\ref{ZW2}):
\begin{eqnarray}
\lim_{|\tau-\tau'|\to\infty}\overline{\langle\psi({\bf x},\tau)\psi^{*}({\bf x},\tau')\rangle}=n_{0}({\bf x})+q({\bf x})\,.
\end{eqnarray}
Note, furthermore, that the localization of the Bose-glass states
can be inferred from the spatial exponential fall-off of the correlation
function $g_{2}({\bf x}-{\bf x'},\frac{{\bf x}+{\bf x'}}{2};\tau-\tau')$
describing correlations of the locally condensed component. In the
Bose-glass phase Eq. \eqref{qg} yields $-\mu+2g\Sigma({\bf x})+V({\bf x})-\frac{D}{\hbar}Q_{0}({\bf x})=\left[D\Gamma\left(2-\frac{n}{2}\right)\,\left(\frac{M}{2\pi\hbar^{2}}\right)^{n/2}\right]^{\frac{2}{4-n}}$.
Inserting this result into the exponential part of function \eqref{ZW2}
allows us to extract for the zero Matsubara mode $m=0$ the temperature-independent
Larkin length $\mathcal{L}=\frac{\hbar}{\sqrt{2M}}\left[D\Gamma\left(2-\frac{n}{2}\right)\,\left(\frac{M}{2\pi\hbar^{2}}\right)^{n/2}\right]^{\frac{1}{n-4}}$,
which is also found in Refs.~\cite{Nattermann,Natterman2,Intro-90,Falco}.
Note that this Larkin length is independent of both the densities
and the interaction strength $g$, since the Hartree-Fock approximation
is an effective free-particle theory.

\end{document}